\documentclass[review,numbers,sort&compress]{elsarticle}

\usepackage{geometry}
\journal{Computers \& Fluids}

\graphicspath{{pic/}}
\date{}
\usepackage{booktabs}
\usepackage{siunitx}
\usepackage{amsmath}
\usepackage{subfigure}
\usepackage{float}
\usepackage[capitalize]{cleveref}
\usepackage{xcolor}
\usepackage{amsfonts}
\usepackage{bm}
\usepackage{diagbox}

\usepackage{caption}
\newcommand{\blueblack}{\color{black}}

\begin{document}

\begin{frontmatter}

\title{Data-driven approach for modeling Reynolds stress tensor with invariance preservation}

\author[mymainaddress,mysecondaryaddress]{Xuepeng Fu}

\author[mymainaddress,mysecondaryaddress]{Shixiao Fu\corref{mycorrespondingauthor}}
\cortext[mycorrespondingauthor]{Corresponding author}
\ead{shixiao.fu@sjtu.edu.cn}

\author[mythirdaddress]{Chang Liu}
\author[mymainaddress,mysecondaryaddress]{Mengmeng Zhang}
\author[myfourthaddress]{Qihan Hu}

\address[mymainaddress]{State Key Laboratory of Ocean Engineering, Shanghai Jiao Tong University, Shanghai, 200240, China}
\address[mysecondaryaddress]{Institute of Polar and Ocean Technology, Institute of Marine Equipment, Shanghai Jiao Tong University, Shanghai, 200240, China}
\address[mythirdaddress]{Department of Mechanical Engineering, University of Connecticut, Storrs, Connecticut 06269, USA}
\address[myfourthaddress]{Center for Biomedical Engineering, School of Information Science and Technology, Fudan University, Shanghai 200433, China}

\begin{abstract}
{\blueblack The present study represents a data-driven turbulent model with Galilean invariance preservation based on machine learning algorithm. The fully connected neural network (FCNN) and tensor basis neural network (TBNN) [Ling et al. (2016)] are established. The models are trained based on five kinds of flow cases with Reynolds Averaged Navier-Stokes (RANS) and high-fidelity data. The mappings between two invariant sets, mean strain rate tensor and mean rotation rate tensor as well as additional consideration of invariants of turbulent kinetic energy gradients, and the Reynolds stress anisotropy tensor are trained. The prediction of the Reynolds stress anisotropy tensor is treated as user's defined RANS turbulent model with a modified turbulent kinetic energy transport equation. The results show that both FCNN and TBNN models can provide more accurate predictions of the anisotropy tensor and turbulent state in square duct flow and periodic flow cases compared to the RANS model. The machine learning based turbulent model with turbulent kinetic energy gradient related invariants can improve the prediction precision compared with only mean strain rate tensor and mean rotation rate tensor based models. The TBNN model is able to predict a better flow velocity profile compared with FCNN model due to a prior physical knowledge.}
\end{abstract}

\begin{keyword}
turbulence model\sep machine learning \sep Reynolds stress anisotropy tensor \sep tensor basis neural network
\end{keyword}

\end{frontmatter}


\section{Introduction}

In most industrial applications where resources and time are limited, the choice is often made to solve the Reynolds averaged Navier-Stokes equations (RANS) in turbulent flow simulations \cite{slotnick2014cfd}. {\blueblack The RANS model employed ensemble/time-averaging method to eliminate temporal dependency for stationary turbulence, which will produce an term named Reynolds stress need turbulence modelling closure.} Turbulence modeling is a primary source of uncertainty in the computational fluid dynamics (CFD) simulations of turbulent flows. {\blueblack Many nonlinear eddy models based on the Boussinesq approximation, such as the $k-\epsilon$ and $k-\omega$ models, and Reynolds stress transport models (RSTM), have been proposed to model Reynolds stress \cite{wilcox1998turbulence}. However, RANS predictions may be inaccurate in some canonical flows due to the poor description of the effects of the Reynolds stresses on the mean flow \cite{craft1996development}. On the other hand, some CFD data with high fidelity, such as direct numerical simulation (DNS) and highly resolved large eddy simulation (LES), have been reported due to the use of high-performance computers and big data techniques like machine learning algorithms.} Research on data-driven turbulent models has been carried out in the last few years to improve the accuracy of turbulence model, in which supervised machine learning (ML) algorithms have been applied mostly in this research field \cite{duraisamy2019turbulence}.

The pioneer research of ML algorithms in fluid mechanic was to develop a partial differential equation (PDE) solution method by directly mapping the properties of the flow field to the velocity field. \citet{milano2002neural} trained a neural network based on velocity fields of DNS data to reconstruct the near wall turbulent flow. \citet{raissi2019physics} develop the physics-informed neural networks (PINN) with modified loss function incorporates PDE residuals. More specifically, several proposed investigations already exist based on different algorithms for establishing an improved Reynolds stress prediction model \cite{xu2022pde}. \citet{tracey2013application} established a Reynolds stress anisotropy prediction model by kernel-based regression algorithms. \citet{duraisamy2015new} applied two types of ML methods, neural networks and Gaussian processes, to model intermittency in transitional turbulence. \citet{zhang2015machine} used a multiscale Gaussian process to model turbulence production in channel flow.  \citet{ling2015evaluation} used random forests to predict regions of high model form uncertainty in RANS results. {\blueblack \citet{zhang2022ensemble} established a turbulent model based on the ensemble Kalman filter method.} In the latter case, uncertainty quantification has been applied to develop predictive models in the absence of data by incorporating stochastic terms that are intended to capture the effect of modeling assumptions \cite{emory2013modeling,xiao2016quantifying,xiao2019quantification}. Reynolds stress is treated as an ellipsoid structure with perturbation and rigid-body rotation in the studies mentioned above . The Euler-angle-based representations of the perturbation is considered for rotation. The predictions in different coordinate systems needs to be transformed. 

Galilean invariance, which states that laws do not alter in different inertial frames of reference, is a fluid mechanics property pertaining to the law of motion. Specifically, no matter how the frame of reference is rotated, reflected, or translated, the pressure and velocity magnitude will remain unchanged. As mentioned above, the Reynolds stress is described as a three-dimensional ellipsoid with Euler-angle-based representations to ensure Galilean invariance \cite{wu2019representation}. Another approach to meet Galilean invariance in turbulent modelling is establishing the mapping based on invaricance obtained from flow field data. \citet{ling2016machine} investigated a machine learning-based turbulent prediction model. The inputs were invariant terms derived from the flow fields. \citet{duraisamy2015new} and \citet{ling2015evaluation} applied feature selection algorithms to optimize the input features. These models established the mapping between invariants and Quantities of Interest (QoIs, e.g. second principal invariant of Reynolds stress anisotropy tensor \cite{ling2016machine}). No representation constrain based on a prior knowledge is applied for modelling. \citet{pope1975more} proposed a complete representation of Reynolds stress anisotropy tensor with invariants set derived from turbulence data based on tensor valued function representation theory \citep{Spencer1997}. \citet{ling2016reynolds} established the neural network to meet the turbulent model proposed by \citet{pope1975more} named tensor based neural network (TBNN). The TBNN embeds tensor functions into the machine learning algorithm for turbulence simulation. In Pope's turbulent model, only mean strain rate tensor and mean rotation rate tensor are taken into consideration which may lead to inaccuracy. {\blueblack Recent studies show that incorporating turbulent kinetic energy gradients, which reflects the historical characteristics of turbulence, into turbulence models can improve the predictive performance of the model \citep{yin2022iterative,wu2018physics,yin2020feature}. However, the turbulent kinetic energy gradients have not been integrated into the TBNN model and the corresponding generalization performance has not been investigated. 

In the present study, the complete invariants sets based on mean strain rate tensor, mean rotation rate tensor and turbulent kinetic energy gradient are included for turbulence modelling based on a fully connected neural network and a tensor basis neural network algorithm. Five different types of flow cases are involved for training process. Two different types of input data of square duct flow and periodic flow are applied for analyzed. This paper is structured as follows. \cref{methods} introduces the methodology of invariant-based turbulent models and the framework of the present study. \cref{results} discusses the predictions of the Reynolds stress anisotropy tensor and flow velocity profile of square duct flow and periodic flow. \cref{conclu} offers conclusions and future research directions. }

\section{Method}\label{methods}
\subsection{Reynolds-averaged Navier-Stokes equation}
{\blueblack 
For an incompressible, constant density and zero gravity flow, the following RANS equations apply:
\begin{align}
\frac{\partial \bar{u}_i}{\partial x_i} & =0, \\
\frac{\partial \bar{u}_i}{\partial t} + \bar{u}_j \frac{\partial \bar{u}_i}{\partial x_j} & =-\frac{1}{\rho} \frac{\partial \bar{p}}{\partial x_i}+\frac{\partial}{\partial x_j}\left(\nu \frac{\partial \bar{u}_i}{\partial x_j}-\overline{u_i^{\prime} u_j^{\prime}}\right),
\end{align} 
where $\bar{u}_i$ is the mean velocity, $\bar{p}$ is the mean pressure and $\nu$ is the kinematic viscosity. The Reynolds stress term $\overline{u_i^{\prime} u_j^{\prime}}$ ($\tau_{ij}$ and  $\boldsymbol{\tau}$) can be expressed as:
\begin{equation}\label{uiujk}
    \tau_{ij} = \overline{u_i^{\prime} u_j^{\prime}}=  \frac{2}{3} k \delta_{i j}+ a_{i j},
\end{equation}
where $k = \frac12 tr(\overline{u_i^{\prime} u_j^{\prime}})$ is the turbulence kinetic energy, $\delta_{i j}$ is the Kronecker delta and $a_{ij}$ is the Reynolds stress anisotropy tensor. The anisotropic part of the Reynolds stresses is important and effective in transporting momentum. The isotropic part is simply absorbed into a modified pressure term.}

Based on the Boussinesq approximation ($a_{ij}=-2\nu_t S_{ij}$), the RANS equations can be written as: 

 \begin{align}
\frac{\partial \bar{u}_i}{\partial x_i} & =0, \\
\frac{\partial \bar{u}_i}{\partial t} + \bar{u}_j \frac{\partial \bar{u}_i}{\partial x_j} & =-\frac{1}{\rho} \frac{\partial \bar{p'}}{\partial x_i} + 2(\nu + \nu_t) \frac{\partial S_{ij}}{\partial x_j},
\end{align} 
where $\bar{p'} = \bar{p}+\frac23 \rho k$, $\nu_t$ is the turbulent viscosity and $S_{ij}$ is the mean strain rate tensor. The turbulent viscosity $\nu_t$ is modeled by PDEs in different turbulent models. For example, the $k-\omega$ turbulent model is modeled by solving the governing equations of $k$ and $\omega$ and $\nu_t=k/ \omega$:
\begin{equation}\label{komegarey}
    \tau_{ij} = \frac{k}{\omega}S_{ij},\;\omega = \frac{\varepsilon}{C_\mu k},
\end{equation}
where $\omega$ is the specific turbulence dissipation rate, and $\varepsilon$ is the turbulence dissipation. $C_{\mu} = 0.09$ is applied in the present study \cite{menter1994two}.

Another class of RANS turbulence models that will briefly be discussed is the Reynolds stress transport model (RSTM). Therein, the Reynolds stress is modeled by the governing equation of Reynolds stress $\overline{u_i^{\prime} u_j^{\prime}}$. RSTM turbulence models are more likely to diverge than Boussinesq approximation based models \cite{launder1975progress}. Consequently, the Boussinesq approximation based models remain the preferred method for many flow cases.

\subsection{Reynolds stress and realizability}
As a positive semi-definite matrix, the eigenvalues of Reynolds stress $\tau_{ij}$ are real and non-negative:
\begin{equation}
    \tau_{ii} \geq 0,\;\; i=1,2,3,
\end{equation}

\noindent Considering Cauchy-Schwarz inequality:
\begin{equation}\label{eqtau}
    \tau_{ij}^2 \leq \tau_{ii}\tau_{jj},\;\; i,j=1,2,3\;(i\neq j).
\end{equation}

\noindent Therefore, the diagonal values of the Reynolds stress are within $[0,2k]$ based on \cref{uiujk}, and the off-diagonal values are within $[-k,k]$. The Reynolds stress tensor can be expressed as:
\begin{equation}
    \tau_{i j}= 2k\left(\frac{1}{3} \delta_{i j}+ b_{i j}\right),
\end{equation}
where $b_{i j}$ (also $\boldsymbol{b}$) is the non-dimensional Reynolds stress anisotropy tensor, which can be expressed as:
\begin{equation}
    b_{i j}=\frac{\tau_{i j}}{2 k}-\frac{1}{3} \delta_{i j}.
\end{equation}

Considering \cref{eqtau}, the diagonal and off-diagonal values of  non-dimensional Reynolds stress anisotropy tensor meet \citep{pope_2000}:
\begin{equation}\label{bijlessbii}
    b_{i j}\leq b_{ii} + b_{jj} +\frac23,\;\; i,j=1,2,3\;(i\neq j).
\end{equation}

\noindent Then, the Reynolds stress tensor can be transformed as follows by eigenvalue decomposition:
\begin{equation}
    \tau_{i j}= 2 k\left(\frac{1}{3} \delta_{i j}+v_{i n} \Lambda_{n l} v_{j l}\right),
\end{equation}
where $v_{ij}$ is the eigenvector and $\Lambda_{i j} = \operatorname{diag}[\lambda_1,\lambda_2,\lambda_3]$ is the diagonal matrix containing the eigenvalues $\lambda_i$ of $b_{ij}$ with $\lambda_{1}+\lambda_{2}+\lambda_{3}=0$ of $b_{ij}$.

The eigenvalues of the Reynolds stress tensor ($\phi_i$) can be related to the eigenvalues of the Reynolds stress anisotropy tensor ($\lambda_i$) by:
\begin{equation}
    \lambda_{i}=\frac{\phi_{i}}{2 k}-\frac{1}{3}.
\end{equation}

\noindent The $\phi_i$ falls in $[0,2k]$, then $\lambda_i$ are in the range $[-\frac13,\frac23]$. The different eigenvalues satisfy the physical boundaries as \cite{banerjee2007presentation}:
\begin{equation}\label{eigenvalueofbij}
\begin{aligned}
\lambda_{1} &\geq\left(3\left|\lambda_{2}\right|-\lambda_{2}\right) / 2 \\
\quad \lambda_{1} &\leq 1 / 3-\lambda_{2}
\end{aligned}
\qquad (\lambda_{1}\geq\lambda_{2}\geq\lambda_{3})
\end{equation}

{\blueblack
Overall, there are the following fundamental constraints of the realizability of the non-dimensional Reynolds stress anisotropy tensor $b_{i j}$: (1) the diagonal value of $b_{ij}$ should be greater than $-\frac13$, and the trace should be 0. (2) The value of $b_{i j}$ should satisfy \cref{bijlessbii}. (3) The eigenvalues of $b_{i j}$ should meet \cref{eigenvalueofbij}. More detailed derivations are referred to \citet{pope_2000} and \citet{banerjee2007presentation}.}

\begin{figure}[htbp!]
	\centering
	\includegraphics[width=.8\textwidth]{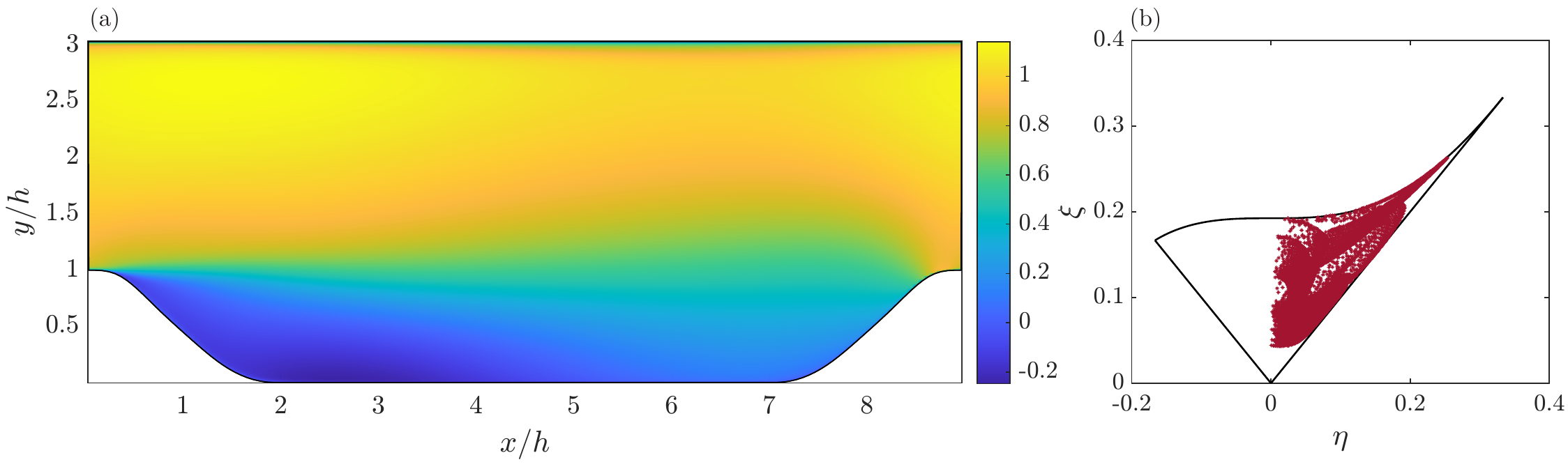}
	\caption{Periodic hills at $Re = 10545$ \cite{breuer2009flow}. (a) mean velocity $U$; (b) anisotropy-invariant map of the Lumley triangle.}
	\label{piclumtri}
\end{figure}

\begin{figure}[htbp!]
	\centering
	\includegraphics[width=.95\textwidth]{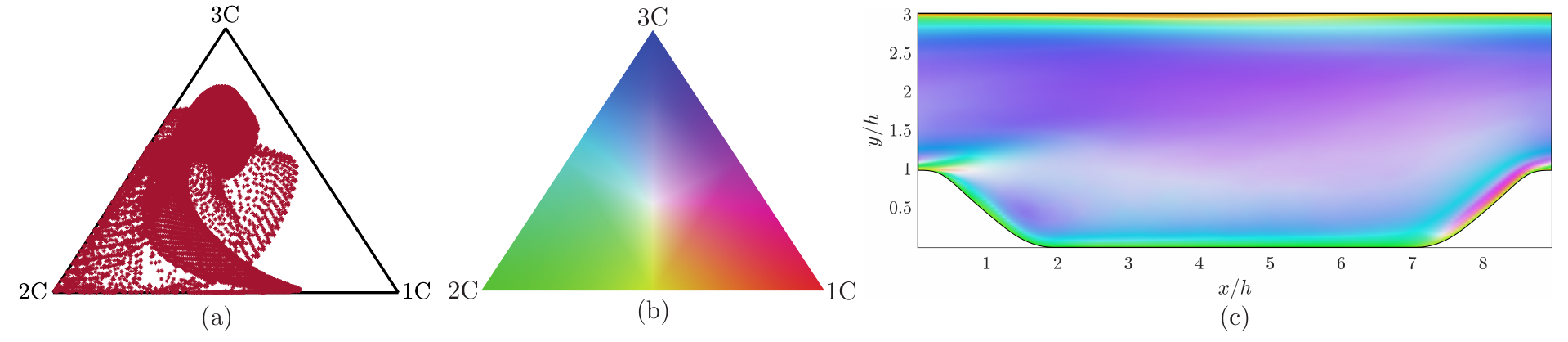}
	\caption{Turbulent states of periodic hills at $Re = 10545$ \cite{breuer2009flow} with Barycentric map. (a) Barycentric map; (b) colored Barycentric map used for colorbar; (c)  anisotropy in the flow field (turbulent state).}
	\label{picbarytri}
\end{figure}

{\blueblack 
There exist three principle invariants of $b_{ij}$: $I_1=b_{i i}$, $I_2=b_{i j} b_{j i}$, and $I_3=b_{i j} b_{i n} b_{j n}$. $I_1$ is the trace of $b_{ij}$ equals to zero.} Therefore, there exist only two independent invariants of the anisotropy tensor $I_2$ and $I_3$. Many realizable states of turbulence anisotropy (noted as anisotropy-invariant maps) have been proposed, such as the Lumley triangle \cite{lumley1977return} and Barycentric map \cite{banerjee2007presentation}. \cref{piclumtri} displayed turbulent states of periodic hills at $Re = 10545$ \cite{breuer2009flow} with a Lumley triangle \cite{lumley1979computational}. \cref{picbarytri} displays the Barycentric map \cite{banerjee2007presentation} of the same flow case. Different from Lumley triangle, an equilateral triangle is used for plotting the physical realizability in the Barycentric map as \cref{picbarytri}(a) shows. We can color the Barycentric map with Red-Green-Blue(RGB) map as colormap shown in \cref{picbarytri}(b), and the flow field can be colored for visualization based colormap of \cref{picbarytri}(b) as \cref{picbarytri}(c) shows. In the present study, the turbulent state is refereed to the anisotropy in the flow field. There exist three corners of Barycentric map refer to three limiting state of turbulence: one-component state (1C), two-component state (2C) and  three-component state (3C). The 1C state means turbulence exists only one component of turbulent kinetic energy, turbulence in an area of this type is only along one direction. The 2C state of turbulence means one component of turbulent kinetic energy vanishes with the remaining two being equal, and 3C state refer to isotropic turbulence.

\subsection{Representation of isotropic tensor-valued function }
The Navier-Stokes equations are Galilean invariant, which means that the equations are the same in all inertial frames of reference, and the model of the Reynolds stress anisotropy tensor should follow this constraint. Otherwise, the predicted results will be different for flow fields with axes defined in different directions.

Define a scalar-valued function $f:\left \{ \mathbb{R}^{3 \times 3},\mathbb{R}^{1 \times 3}\right\}\rightarrow \mathbb{R}$. The necessary and sufficient condition that $f$ is Galilean invariant applied when:
\begin{equation}
    f\left( \boldsymbol{Q} \boldsymbol { A } _ { 1 } \boldsymbol{Q}^{T}, \boldsymbol{Q} \boldsymbol{ A } _ { 2 } \boldsymbol{Q}^{T}, \ldots, \boldsymbol{Q} \mathbf { A } _ { n } \boldsymbol{Q}^{T},\boldsymbol{Q}\boldsymbol{v}\right)=f\left(\boldsymbol{A}_{1}, \boldsymbol{A}_{2}, \ldots, \boldsymbol{A}_{n} \ldots, \boldsymbol{v} \right), \; \boldsymbol{A}_i \in \mathbb{R}^{3 \times 3},\;\boldsymbol{v} \in \mathbb{R}^{1 \times 3},\;\boldsymbol{Q} \in O(3),
\end{equation}
where $O(3)$ is the orthogonal group, and the function $f$ is called as isotropic function. Furthermore, an isotropic tensor-valued function $\boldsymbol{G}:\mathbb{R}^{3 \times 3} \rightarrow \mathbb{R}^{3 \times 3}$ with one variable $\boldsymbol { A }$ can be spectrally decomposed as:
\begin{equation}\label{tenvf}
    \boldsymbol{G}(\boldsymbol{A})=\sum_{i=1}^{n} t_{i}\left(\lambda_{1}, \lambda_{2}, \lambda_{3}\right) \boldsymbol{E}_{i},
\end{equation}
where $t_i:\mathbb{R}^3 \rightarrow \mathbb{R}$ is a scalar function of eigenvalues of $\boldsymbol{A}$, $\lambda_i$ are the invariants of $\boldsymbol{A}$, and $\boldsymbol{E}_{i}$ is the tensor basis.

As per the Cayley-Hamilton theorem \cite{spencer1958theory}, there exists a limited number of linearly independent tensors bases (form-invariants) that can be formed from $\boldsymbol{A}$. It can be derived from the polynomial expression of $\boldsymbol{A}$ with three tensor bases as $\boldsymbol{I}$ (identity tensor), $\boldsymbol{A}$ and $\boldsymbol{A^2}$. Then \cref{tenvf} can be further expressed as:
\begin{equation}
    \boldsymbol{G}(\boldsymbol{A})=\sum_{i=1}^{3} t_{i}\left(\lambda_{1}, \lambda_{2}, \lambda_{3}\right) \boldsymbol{G}_{i}, \quad \boldsymbol{G}_i = \boldsymbol{A} ^{i-1}.
\end{equation}

In the present study, the normalized mean strain rate tensor and mean rotation rate tensor are defined as:
\begin{equation}
\begin{aligned}
    \boldsymbol{S} &= \hat{S}_{i j}= \frac{{S}_{i j}}{(|\alpha|+|S_{i j}|)},\;{S}_{i j} = \frac{1}{2}\left(\frac{\partial \bar{u}_{i}}{\partial x_{j}}+\frac{\partial \bar{u}_{j}}{\partial x_{i}}\right),\\
    \boldsymbol{R} &= \hat{R}_{i j}= \frac{{R}_{i j}}{(|\alpha|+|R_{i j}|)},\;{R}_{i j} = \frac{1}{2}\left(\frac{\partial \bar{u}_{i}}{\partial x_{j}}-\frac{\partial \bar{u}_{j}}{\partial x_{i}}\right),\\
\end{aligned}    
\end{equation}
where $\alpha = \varepsilon/k$ is the normalization factor. The normalized mean strain rate tensor is expressed as $S_{ij}/\alpha$ in the previous research \cite{pope1975more,ling2016reynolds}. The normalization method in the present study will allow a smaller range of values.

The traditional turbulent model based on the Boussinesq approximation assumes that the Reynolds stress anisotropy tensor is related to the mean strain rate tensor $\boldsymbol{b}\sim \boldsymbol{S}$. If the mean rotation rate tensor is added to the independent variable, a tensor-valued function $\boldsymbol{b} = \boldsymbol{F}(\boldsymbol{S},\boldsymbol{R})$ is adequately modeled.

Wang and Smith have investigated the representation theorem for isotropic functions \cite{wang1970new,smith1971isotropic}. Eringen and Zheng summarized complete and irreducible invariants and tensors basis of second-order symmetric tensors, asymmetric tensors and vectors \cite{eringen1980mechanics,zheng1994theory}. For the second-order symmetric tensor $A$ and asymmetric tensor $R$, the complete and irreducible set contains 7 invariants and 10 tensor bases. The invariants can be expressed as:
\begin{equation}
    \tilde{\lambda}_1  = \operatorname{tr}(\boldsymbol{S}),\; \tilde{\lambda}_2 = \operatorname{tr}(\boldsymbol{S}^2),\; \tilde{\lambda}_3 = \operatorname{tr}(\boldsymbol{S}^3),\; \tilde{\lambda}_4 = \operatorname{tr}(\boldsymbol{R}^2),\; \tilde{\lambda}_5 = \operatorname{tr}(\boldsymbol{S}\boldsymbol{R}^2),\; \tilde{\lambda}_6 = \operatorname{tr}(\boldsymbol{S}^2\boldsymbol{R}^2),\; \tilde{\lambda}_7 = \operatorname{tr}(\boldsymbol{S}^2\boldsymbol{R}^2\boldsymbol{S}\boldsymbol{R})
\end{equation}

In the 7 invariants, $\tilde{\lambda}_1 = 0$ due to the continuity equation of incompressible fluids, and $\tilde{\lambda}_7^2$ has an implicit function connection with the previous six invariants $\tilde{\lambda}_7^2 = t(\tilde{\lambda}_1,...,\tilde{\lambda}_6)$, with the proof is displayed in Appendix A. Therefore, the minima set contains five independent invariants. The tensor-valued function $\boldsymbol{F}:\mathbb{R}^{3 \times 3} \rightarrow \mathbb{R}^{3 \times 3}$ can be expressed as:
\begin{equation}\label{popeb}
    \boldsymbol{b} = \sum_{i=1}^{10} g_i(\lambda_1,...,\lambda_5)\boldsymbol{T}^i,
\end{equation}
where:
\begin{equation}\label{T1toT10}
\left\{ \begin{array}{lrl}
\boldsymbol{T}^{1} =\boldsymbol{S}, && \boldsymbol{T}^{6}=\boldsymbol{R}^{2} \boldsymbol{S}+\boldsymbol{S R}^{2}-\frac{2}{3} \boldsymbol{I} \cdot \operatorname{tr}\left(\boldsymbol{S} \boldsymbol{R}^{2}\right), \\
\boldsymbol{T}^{2} =\boldsymbol{S} \boldsymbol{R}-\boldsymbol{R} \boldsymbol{S}, & & \boldsymbol{T}^{7}=\boldsymbol{R} \boldsymbol{R}^{2}-\boldsymbol{R}^{2} \boldsymbol{S} \boldsymbol{R}, \\
\boldsymbol{T}^{3} =\boldsymbol{S}^{2}-\frac{1}{3} \boldsymbol{I} \cdot \operatorname{tr}\left(\boldsymbol{S}^{2}\right), & & \boldsymbol{T}^{8}=\boldsymbol{S} \boldsymbol{R} \boldsymbol{S}^{2}-\boldsymbol{S}^{2} \boldsymbol{R S}, \\
\boldsymbol{T}^{4} =\boldsymbol{R}^{2}-\frac{1}{3} \boldsymbol{I} \cdot \operatorname{tr}\left(\boldsymbol{R}^{2}\right), \qquad& & \boldsymbol{T}^{9}=\boldsymbol{R}^{2} \boldsymbol{S}^{2}+\boldsymbol{S}^{2} \boldsymbol{R}^{2}-\frac{2}{3} \boldsymbol{I} \cdot \operatorname{tr}\left(\boldsymbol{S}^{2} \boldsymbol{R}^{2}\right), \\
\boldsymbol{T}^{5} =\boldsymbol{R} \boldsymbol{S}^{2}-\boldsymbol{S}^{2} \boldsymbol{R}, & & \boldsymbol{T}^{10}=\boldsymbol{R S}^{2} \boldsymbol{R}^{2}-\boldsymbol{R}^{2} \boldsymbol{S}^{2} \boldsymbol{R}, \\
\end{array}\right.
\end{equation}

\begin{equation}\label{lambda1to5}
    \lambda_{1}=\operatorname{tr}\left(\boldsymbol{S}^{2}\right), \quad \lambda_{2}=\operatorname{tr}\left(\boldsymbol{R}^{2}\right), \quad \lambda_{3}=\operatorname{tr}\left(\boldsymbol{S}^{3}\right), \quad \lambda_{4}=\operatorname{tr}\left(\boldsymbol{R}^{2} \boldsymbol{S}\right), \quad \lambda_{5}=\operatorname{tr}\left(\boldsymbol{R}^{2} \boldsymbol{S}^{2}\right).
\end{equation}

\noindent This turbulent model was first proposed by Pope \cite{pope1975more} and the same notation is applied in the present study.

\subsection{Tensor basis neural network}
\begin{figure}[htbp!]
	\centering
	\includegraphics[width=.55\textwidth]{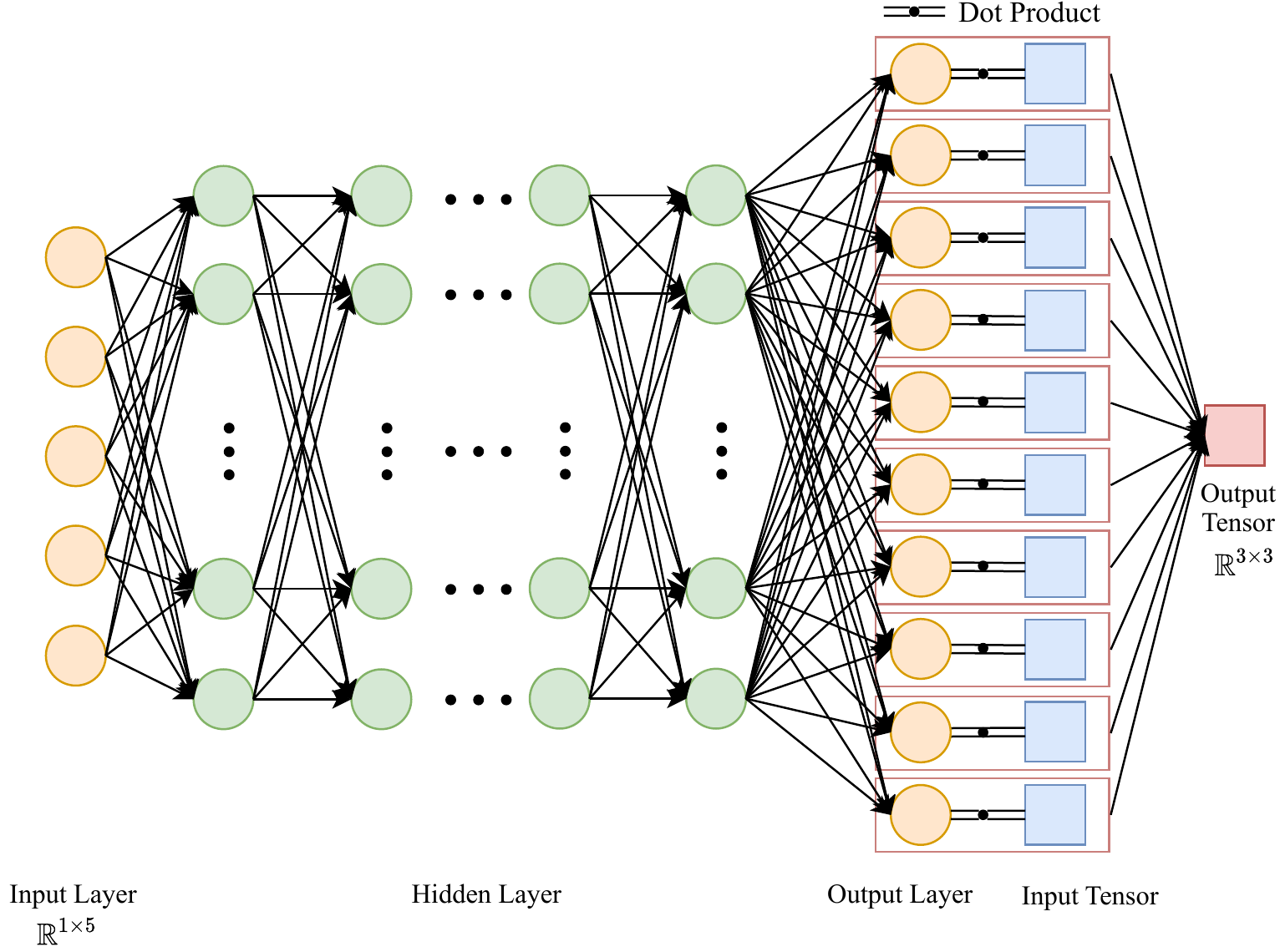}
	\caption{Schematic of TBNN-5. A fully connected neural network contains an input layer (five invariants based on $\boldsymbol{S}$ and $\boldsymbol{R}$), hidden layer and output layer (ten scalar function values in \cref{popeb}). The output tensor is the sum of the dot product of the output layer values and ten input tensors (ten tensor bases based on $\boldsymbol{S}$ and $\boldsymbol{R}$).}
	\label{pictnbbfig}
\end{figure}

\citet{ling2016reynolds} developed a neural network architecture named the tensor basis neural network (TBNN) to implement Pope's turbulent model. \cref{pictnbbfig} represents the schematic of TBNN-5. The TBNN-5 means TBNN with five input features which is the same with Pope's turbulent model. A fully connected neural network with five input neurons and ten output neurons produces $g_i (1\leq i \leq 10)$. The output layer values then are applied as dot product with the 10 input tensors. The loss function is the mean square error (MSE) of the output tensor. 

{\blueblack
In the $k-\omega$ turbulent model, the Reynolds stress tensor is modeled by \cref{komegarey}, which means that the Reynolds stress anisotropy tensor of the RANS model is also a deterministic function of $\boldsymbol{S}$. In the present study, the discrepancy of $\Delta\boldsymbol{b}$ between the high-fidelity data and RANS data will slightly reduce data fluctuations such as the scalar function $g_{i}$ (see Appendix B). Therefore, a tensor-valued function $\boldsymbol{F}_1:\mathbb{R}^{3 \times 3} \rightarrow \mathbb{R}^{3 \times 3}$ is trained by the neural network the same as the TBNN in the present study: }

\begin{equation}\label{popeb1}
    \Delta\boldsymbol{b} = \sum_{i=1}^{10} g_i(\lambda_1,...,\lambda_5)\boldsymbol{T}^i.
\end{equation}

\begin{figure}[htbp!]
	\centering
	\includegraphics[width=.45\textwidth]{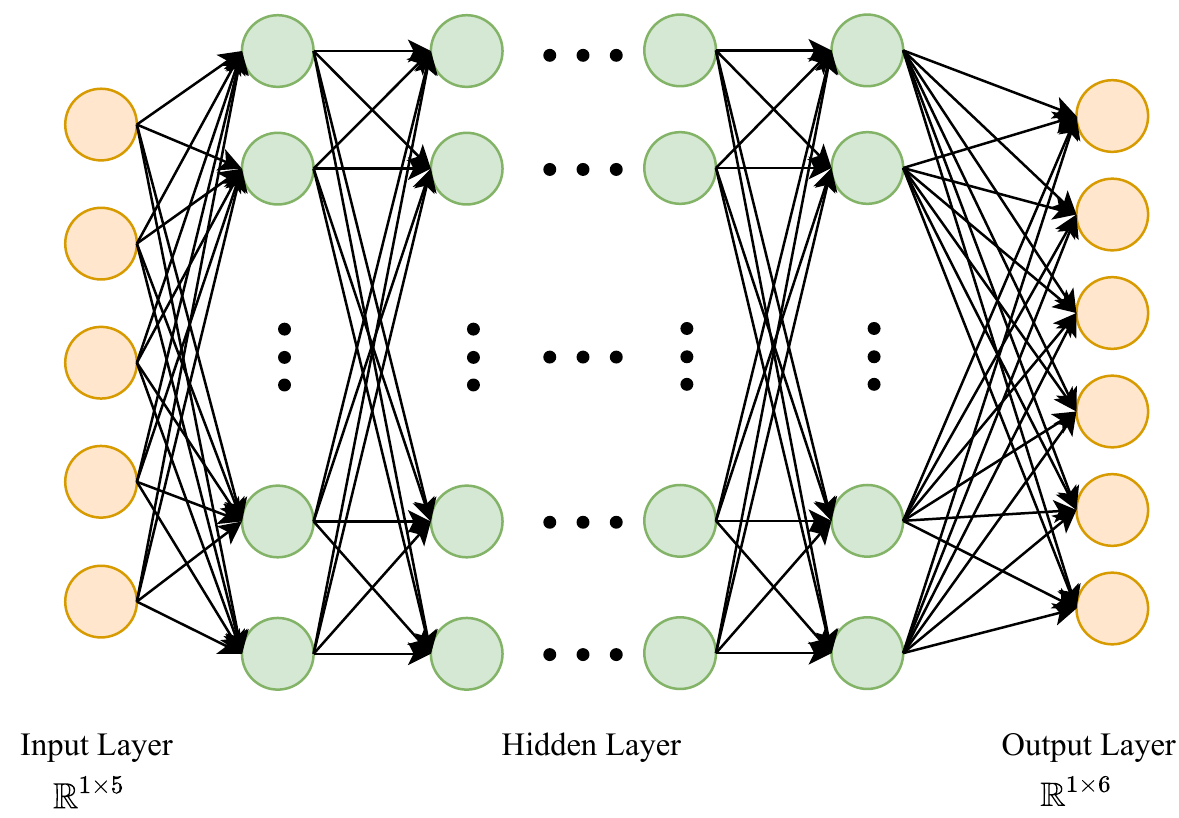}
	\caption{Schematic of FCNN-5 model. A fully connected neural network contains an input layer (five invariants based on $\boldsymbol{S}$ and $\boldsymbol{R}$), hidden layer and output layer (six independent values of $\Delta\boldsymbol{b}$).}
	\label{picfcnn1fig}
\end{figure}

\noindent The neural network structure is the same as \cref{pictnbbfig} shows, and the values of $g_i$ and $\boldsymbol{T}^i$ are the same as \cref{T1toT10} and \cref{lambda1to5} display. Meanwhile, a fully connected neural network (FCNN) model is also trained to obtain regression function $\boldsymbol{F}$ as \cref{picfcnn1fig} shows. The input layer contains the five invariants, and the output layer is the six independent values of $\Delta\boldsymbol{b}$ .

 \begin{figure}[htbp!]
	\centering
	\includegraphics[width=.9\textwidth]{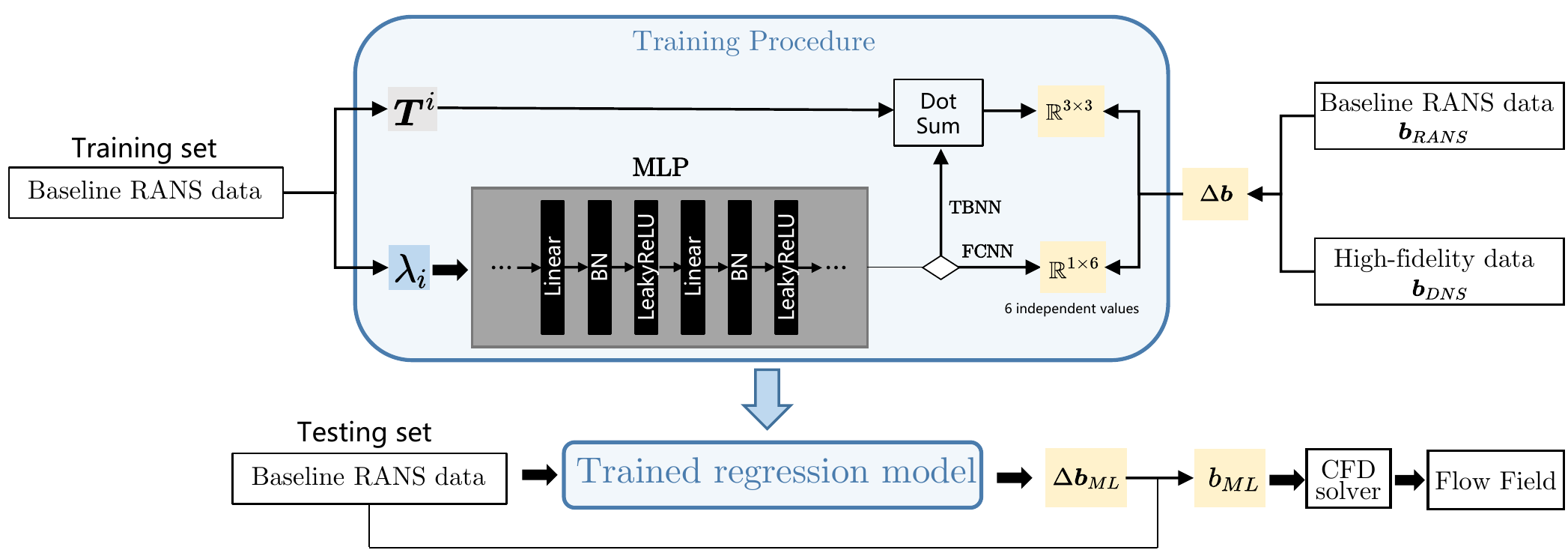}
	\caption{Framework of the present study. $\boldsymbol{T}^i$ and $\lambda_i$ are the tensor basis and invariants obtained from the training set RANS flow case; the multilayer perceptron (MLP) with linear, batch normalization (BN) and LeakyReLU activator layers are established; the TBNN method applies a dot product with the MLP output layer ($\mathbb{R}^{1\times 10}$) with corresponding $\boldsymbol{T}^i$ and summary; the FCNN method will directly output $\Delta\boldsymbol{b}$.}
	\label{picprocedure}
\end{figure} 

{\blueblack
The mean strain rate tensor and mean rotation rate tensor are considered in the Pope turbulent model \citep{pope1975more}. In the present study, we account for the turbulent kinetic energy gradient reflecting historical characteristics of turbulence \citep{yin2022iterative,yin2020feature,wu2018physics} in turbulent modeling, which includes establishing a tensor function as follows:

\begin{equation}
    \Delta\boldsymbol{b} = \boldsymbol{F_2}(\boldsymbol{S},\boldsymbol{R},\widehat{\nabla k}),
\end{equation}
where $\widehat{\nabla k}$ is the non-dimensional ${\nabla k}$ by non-dimensional coefficient $\frac{\epsilon}{\sqrt{k}}$ \citep{wu2018physics}:
\begin{equation}
    \widehat{\nabla k} = \dfrac{\nabla k}{(|\frac{\epsilon}{\sqrt{k}}|+|\nabla k|)}.
\end{equation}
\noindent The complete invariants set based on $S$, $R$ and $\widehat{\nabla k}$ are as follows: \cite{eringen1980mechanics,zheng1994theory}:
\begin{equation}\label{T1toT10V2}
\left\{ \begin{array}{lrl}
\boldsymbol{T}^{1} =\boldsymbol{S}, && \boldsymbol{T}^{6}=\boldsymbol{R}^{2} \boldsymbol{S}+\boldsymbol{S R}^{2}-\frac{2}{3} \boldsymbol{I} \cdot \operatorname{tr}\left(\boldsymbol{S} \boldsymbol{R}^{2}\right), \\
\boldsymbol{T}^{2} =\boldsymbol{S} \boldsymbol{R}-\boldsymbol{R} \boldsymbol{S}, & & \boldsymbol{T}^{7}=\boldsymbol{R} \boldsymbol{R}^{2}-\boldsymbol{R}^{2} \boldsymbol{S} \boldsymbol{R}, \\
\boldsymbol{T}^{3} =\boldsymbol{S}^{2}-\frac{1}{3} \boldsymbol{I} \cdot \operatorname{tr}\left(\boldsymbol{S}^{2}\right), & & \boldsymbol{T}^{8}=\boldsymbol{S} \boldsymbol{R} \boldsymbol{S}^{2}-\boldsymbol{S}^{2} \boldsymbol{R S}, \\
\boldsymbol{T}^{4} =\boldsymbol{R}^{2}-\frac{1}{3} \boldsymbol{I} \cdot \operatorname{tr}\left(\boldsymbol{R}^{2}\right), \qquad& & \boldsymbol{T}^{9}=\boldsymbol{R}^{2} \boldsymbol{S}^{2}+\boldsymbol{S}^{2} \boldsymbol{R}^{2}-\frac{2}{3} \boldsymbol{I} \cdot \operatorname{tr}\left(\boldsymbol{S}^{2} \boldsymbol{R}^{2}\right), \\
\boldsymbol{T}^{5} =\boldsymbol{R} \boldsymbol{S}^{2}-\boldsymbol{S}^{2} \boldsymbol{R}, & & \boldsymbol{T}^{10}=\boldsymbol{R S}^{2} \boldsymbol{R}^{2}-\boldsymbol{R}^{2} \boldsymbol{S}^{2} \boldsymbol{R}, \\
\end{array}\right.
\end{equation}

\begin{equation}
\begin{aligned}
      \lambda_{1}&=\operatorname{tr}\left(\boldsymbol{S}^{2}\right), \quad \lambda_{2}=\operatorname{tr}\left(\boldsymbol{R}^{2}\right), \quad \lambda_{3}=\operatorname{tr}\left(\boldsymbol{S}^{3}\right), \quad \lambda_{4}=\operatorname{tr}\left(\boldsymbol{R}^{2} \boldsymbol{S}\right),\\
       \lambda_{5} &=\operatorname{tr}\left(\boldsymbol{R}^{2}\boldsymbol{S}^{2}\right), \quad
      \lambda_{6} = \widehat{\nabla k} \cdot \boldsymbol{S}\widehat{\nabla k}, \quad
    \lambda_{7} = \widehat{\nabla k} \cdot \boldsymbol{S}^2\widehat{\nabla k}, \quad
    \lambda_{8} = \widehat{\nabla k} \cdot \boldsymbol{R}^2\widehat{\nabla k},\\
    \lambda_{9} &= \widehat{\nabla k} \cdot \boldsymbol{S}\boldsymbol{R}\widehat{\nabla k}, \quad
    \lambda_{10} = \widehat{\nabla k} \cdot \boldsymbol{S}^2\boldsymbol{R}\widehat{\nabla k}, \quad
    \lambda_{11} = \widehat{\nabla k} \cdot \boldsymbol{R}\boldsymbol{S}\boldsymbol{R}^2\widehat{\nabla k}, \quad
    \lambda_{12} = \widehat{\nabla k} \cdot \widehat{\nabla k}.
\end{aligned}
\end{equation}
 }
 
It should be noted that the aforementioned invariant set is complete and irreducible, but it is not a minima set. The training of neural networks may be impacted by the presence of implicit function relationships between invariants, but it has no significant impact in the present study. The research of irreducible complete minima invariants set is planed.

\subsection{Propagation of the predicted anisotropy tensor}
The C++ based open source CFD toolbox OpenFOAM is used to simulate the RANS flow field. The $k-\omega$ turbulent model is used in the present study. In the present study, we basically follow this propagation procedure with an underrelaxing coefficient to enhance the conditioning \cite{ling2016reynolds}. The Reynolds stress $\boldsymbol{\tau}$ is treated as:
\begin{equation}
    \boldsymbol{\tau} = \frac23 k\boldsymbol{I} + 2k \boldsymbol{b}_{ML},
\end{equation}
where $\boldsymbol{b}_{ML}$ is obtained from the ML model, and $k$ is calculated by the kinetic energy governing equation with the modified production term $-\boldsymbol{\tau}:\nabla \boldsymbol{\bar{u}}$.

The procedure of the present study is summarized as follows:
\begin{enumerate}
    \item Collect the high-fidelity LES/DNS/experimental turbulence data from the turbulence database. Simulate the baseline flow field by OpenFOAM based on $k-\omega$ RANS turbulent model.
    \item Calculate the tensor basis $\boldsymbol{T}^{n}$ and tensor invariants $\lambda_i$ based on the RANS results. Calculate the Reynolds stress anisotropy tensor $b_{ij}$ based on high-fidelity turbulent data.
    \item Training the regression function $\boldsymbol{F}:\{\boldsymbol{T}^{n},\lambda_i\}\mapsto \Delta\boldsymbol{b}$ and $\boldsymbol{F}:\lambda_i\mapsto \Delta\boldsymbol{b}$ based on data prepared in the previous step and machine learning algorithm.
    \item Propagate the predicted anisotropy tensor into the SIMPLE algorithm with a modified $k$ production term, and the new RANS flow field is simulated.
\end{enumerate}

\cref{picprocedure} represents the training framework of the present study. The flow case data will be divided into a training set and a testing set. The models trained in the present study are the FCNN-5, FCNN-12, TBNN-5 and TBNN-12 models. The FCNN-5 model establishes the regression function of five invariants based on $\boldsymbol{S}$ and $\boldsymbol{R}$ with an output of $\Delta \boldsymbol{b}$. The FCNN-12 is based on the 12 invariants based on $\boldsymbol{S}$, $\boldsymbol{R}$ and $\widehat{\nabla k}$. The output of the FCNN model is $\Delta \boldsymbol{b} \in \mathbb{R}^{1\times 6}$. Compared with the FCNN model, the TBNN model will produce the scalar function values $g_{i} \in \mathbb{R}^{1\times 10}$ and dot product $g_{i}$ with tensor basis $T^{n}\in\mathbb{R}^{3\times3\times10}$, and the output tensor is the sum of the products in the third dimension. More specifically, the input layer of FCNN-5 model is five invariants of $\boldsymbol{S}$, $\boldsymbol{R}$ and that of FCNN-12 model is the 12 invariants based on $\boldsymbol{S}$, $\boldsymbol{R}$ and $\widehat{\nabla k}$.

\section{Results and discussion}\label{results}
\subsection{Turbulence datasets}
\begin{figure}[h]
	\centering
	\includegraphics[width=.85\textwidth]{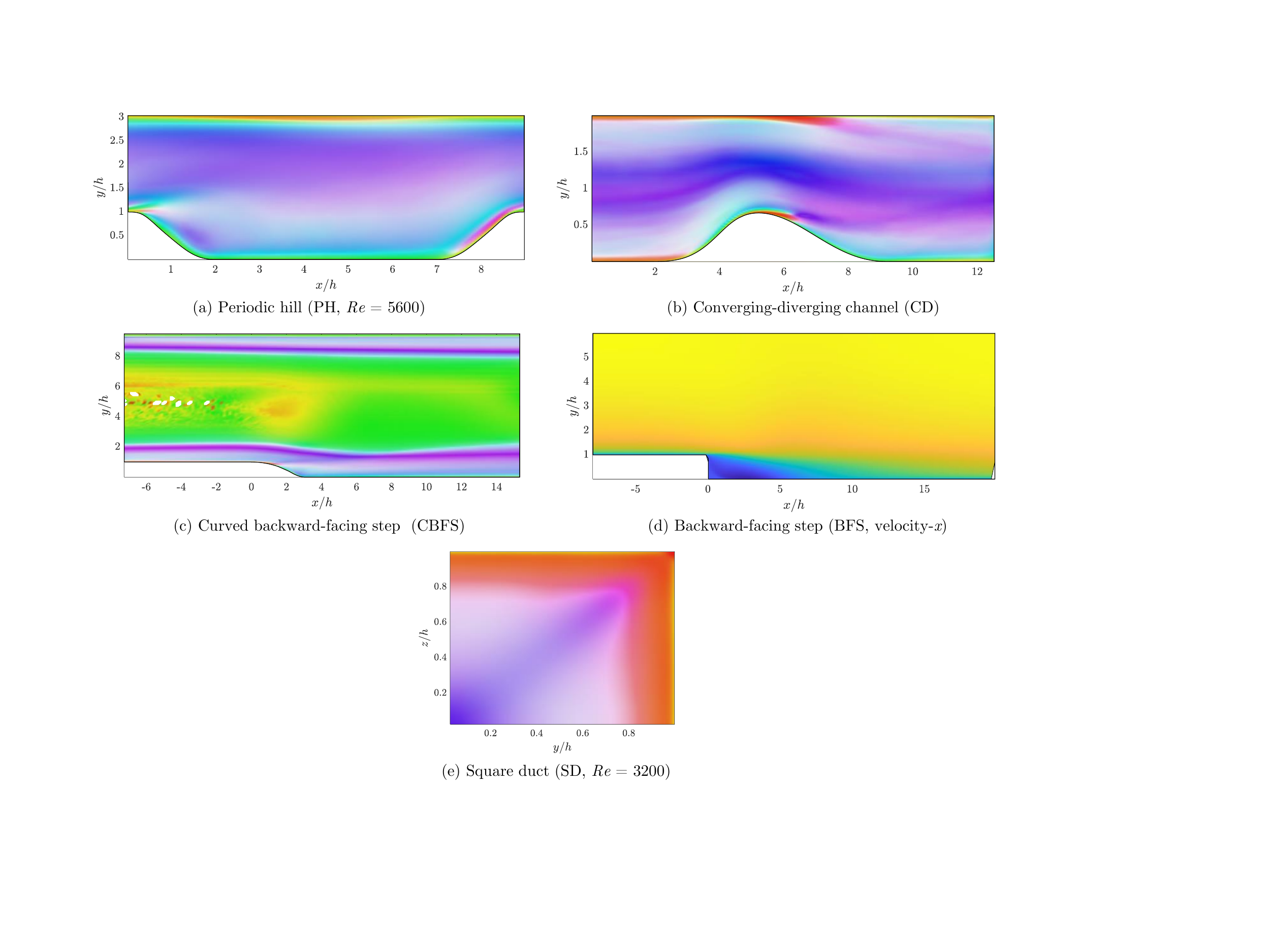}
	\caption{Flow cases used in the present study. The BFS flow case is displayed through velocity in the $x$ direction (blue represents negative value, yellow represents positive value), and the other flow cases are displayed through the turbulent state. The SD flow case only presents the upper right quadrant of the duct, where the flow in the duct removes-of-plane.}
	\label{picflowcase}
\end{figure}

In the present study, five flow cases are involved in machine learning procedures. The DNS and highly resolved LES data are available. The corresponding RANS data are obtained by the $k-\omega$ turbulent model for the Reynolds stress \citep{kaandorp2020data,kaandorpmaster}. \cref{picflowcase} displays the flow cases used in the present study. Five types of relevant flow cases are:
\begin{itemize}
    \item Periodic hills (PH): The data of periodic hill flow fields are obtained from \citet{breuer2009flow}. The Reynolds number $Re$ ranges from $700$ to $10595$ $(Re = 700, 1400, 2800, 5600, 10595)$. Five cases are included in the present study.
    \item Converging-diverging channel (CD): The DNS data of the converging-diverging channel flow field is obtained from \citet{marquillie2011instability} at $Re = 13700$. Public download access to the data is available\footnote{\url{https://turbmodels.larc.nasa.gov/Other_DNS_Data/conv-div-channel12600.html}}.
    \item Curved backward-facing step (CBFS): The LES data of curved backward-facing step flow field is obtained from \citet{bentaleb2012large} and the Reynolds number $Re$ is 12600\footnote{\url{https://turbmodels.larc.nasa.gov/Other_LES_Data/curvedstep.html}}.
    \item Backward-facing step (BFS): The DNS data of backward-facing step flow field is obtained from \citet{le1997direct} at $Re = 5100$
    \footnote{\url{http://cfd.mace.manchester.ac.uk/ercoftac/doku.php?id=cases:case031&s[]=backward&s[]=facing&s[]=step}}.
    \item Square duct (SD): The square duct flow field contains data from eight flow fields with $Re$ ranging from $1800$ to $3500$ from \citet{pinelli2010reynolds}. The Reynolds numbers $Re$ are 1800, 2000, 2200, 2400, 2600, 2900, 3200 and 3500.
\end{itemize}
 
{\blueblack 
In the present study, the testing set consists of PH ($Re = 5600$) and SD ($Re = 3500$), and the remaining data are the training and validation data of the present study. The MLP model contains 16 hidden layers with a combination of 15, 50, 50, 150, 150, 150, 300, 300, 300, 300, 300, 150, 150, 150, 50, and 15 neurons per layer. The structure are determine based on \citet{parashar2020modeling} with parameter optimization. Fine-tuning the network structure parameters has little impact on the prediction results. Kaiming initialization method \citep{he2015delving} is applied to the linear layer. LeakyReLU and batch normalization are used for the activator. The Adam optimizer \citep{kingma2014adam}, with the initial learning rate $1.0\times 10^{-6}$ and a decay rate of $0.999$ for TBNN models and $5.0\times 10^{-7}$ for FCNN models, are applied for training. To avoid overfitting, early stopping is applied in the training process. The incorporation of batch normalization layers has shown a beneficial impact on the prediction results. \cref{piclrandloss} displays the selected learning rate curve and loss curve in TBNN case. The learning rate in training process is gradually decrease.
Mean Square Error (MSE) is applied for loss function, and early stopping is applied in the training process to prevent overfitting when loss function value $L$ does not decrease in $800$ epochs. It can be observed that there is no significant overfitting during the training.

\begin{figure}[!htbp]
	\centering
	\includegraphics[width=.6\textwidth]{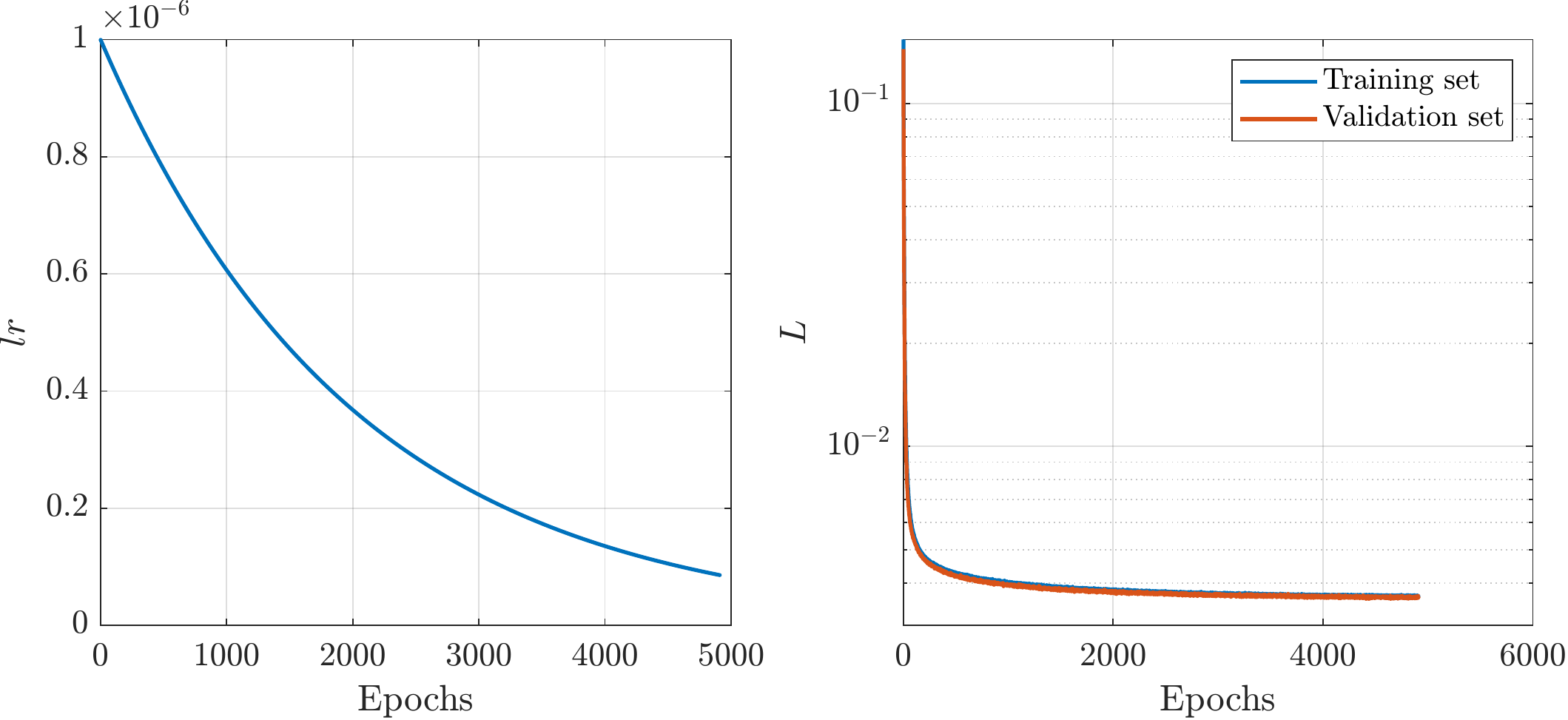}
	\caption{Variation learning rate and loss function with training epochs. Left: learning rates $lr$; right: loss function $L$ in training and validation set. Mean Square Error (MSE) is applied for loss function.}
	\label{piclrandloss}
\end{figure}

}

\subsection{Reynolds stress anisotropy tensor $b_{ij}$ predictions}
\begin{figure}[!htbp]
	\centering
	\includegraphics[width=.95\textwidth]{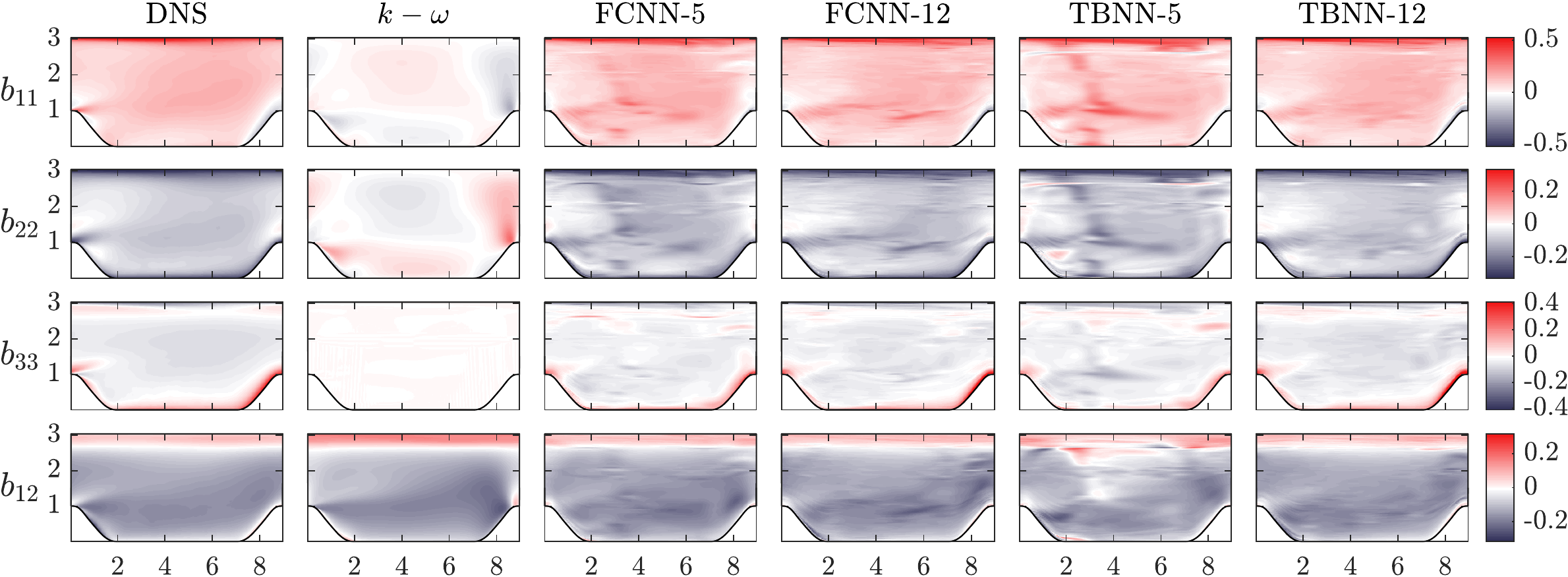}
	\caption{$b_{ij}$ prediction values of PH at $Re = 5600$. $k-\omega$ data is the RANS turbulent model prediction values.}
	\label{picphbij}
\end{figure}

In this section, the Reynolds stress anisotropy tensors $b_{ij}$ of the test set are displayed. \cref{picphbij} displays the contour plot of the $b_{ij}$ prediction values of the PH case at $Re = 5600$. The DNS data are obtained from \citet{breuer2009flow}. The $b_{ij}$ of $k-\omega$ turbulent model data is derived from \cref{komegarey}. The input data of the FCNN and TBNN models are normalized by the mean and std values of the training dataset. Since the flow is homogeneous in the $z$ direction, the $b_{13}$ and $b_{23}$ components are zero \citep{kaandorp2020data}. Here, we represent the other 4 independent values of $b_{ij}$. 

{\blueblack 
It can be noticed that the $k-\omega$ turbulent model can only predict the $b_{12}$ component with acceptable accuracy. We can see that the machine learning turbulent model can not only predict the $b_{12}$ values but the rest components. The root mean square error (RMSE) values of different prediction values versus DNS values are displayed in \cref{RMSEbijSD}. The machine learning based turbulent models yield much better results than traditional turbulence models based on the Boussinesq hypothesis. While the prediction values of machine learning models show higher accuracy but poor continuity. This could be attributed to the size of the dataset used for training and also demonstrates that machine learning turbulence models might exhibit some bias in generalization predictions. The machine learning algorithm searches for a mapping relationship between input and output rather than a predetermined function. The poor continuity also has been found in related studies, a postprocess procedures like Gaussian filter are applied to improve the prediction results and some data smoothing methods based on prior knowledge are applied to obtain better continuity for machine learning turbulence model prediction \citep{kaandorp2020data,wu2019reynolds}. In the present study, the direct prediction is displayed with no filtering since the postprocess procedures have not been investigated thoroughly. Meanwhile, MLP model is a pointwise constrained neural network structure that does not account for the spatial distribution of data, leading to the loss of neighborhood information. Some studies, such as \citet{fang2021high}, attempt to enhance the predictive capabilities of PINN models by integrating convolutional neural network (CNN) models to incorporate neighborhood influence and reduce function space. Related research will be conducted further.}

{\blueblack 
Both the FCNN and TBNN models can predict more accurate values with involement of $\nabla k$ in the present study, as shown in \cref{RMSEbijPH}. In the FCNN model, a direct mapping between the invariants of fluid data and the Reynolds stress anisotropy tensor is established. The TBNN model adds a restriction of the scalar function $g_i$ with a certain function structure, which decreases the overall neural network's capacity, similar to the PINN model \citep{raissi2019physics}. Therefore, the predictions of the TBNN model exhibit comparable or even greater errors of the FCNN model but yield smoother results.}

\begin{table}[!htbp]
\centering
\caption{$b_{ij}$ prediction values of PH at $Re = 5600$. $k-\omega$ data is the RANS turbulent model prediction values.}
\label{RMSEbijPH}
\resizebox{.55\textwidth}{!}{%
\begin{tabular}{cccccc}
\hline
 & \multicolumn{5}{c}{Model} \\ \cline{2-6} 
RMSE Value & RANS & FCNN-5 & FCNN-12 & TBNN-5 & TBNN-12 \\ \hline
$b_{11}$ & 0.1975 & 0.0646 & 0.0611 & 0.0811 & 0.0531 \\
$b_{22}$ & 0.1743 & 0.0428 & 0.0318 & 0.0638 & 0.0251 \\
$b_{33}$ & 0.0948 & 0.0504 & 0.0516 & 0.0667 & 0.0452 \\ 
$b_{12}$ & 0.0658 & 0.0384 & 0.0321 & 0.0483 & 0.0253 \\\hline
$b_{ij}$ & 0.0984 & 0.0357 & 0.0324 & 0.0469 & 0.0307 \\ \hline
\end{tabular}%
}
\end{table}

\begin{figure}[htbp]
	\centering
	\includegraphics[width=.7\textwidth]{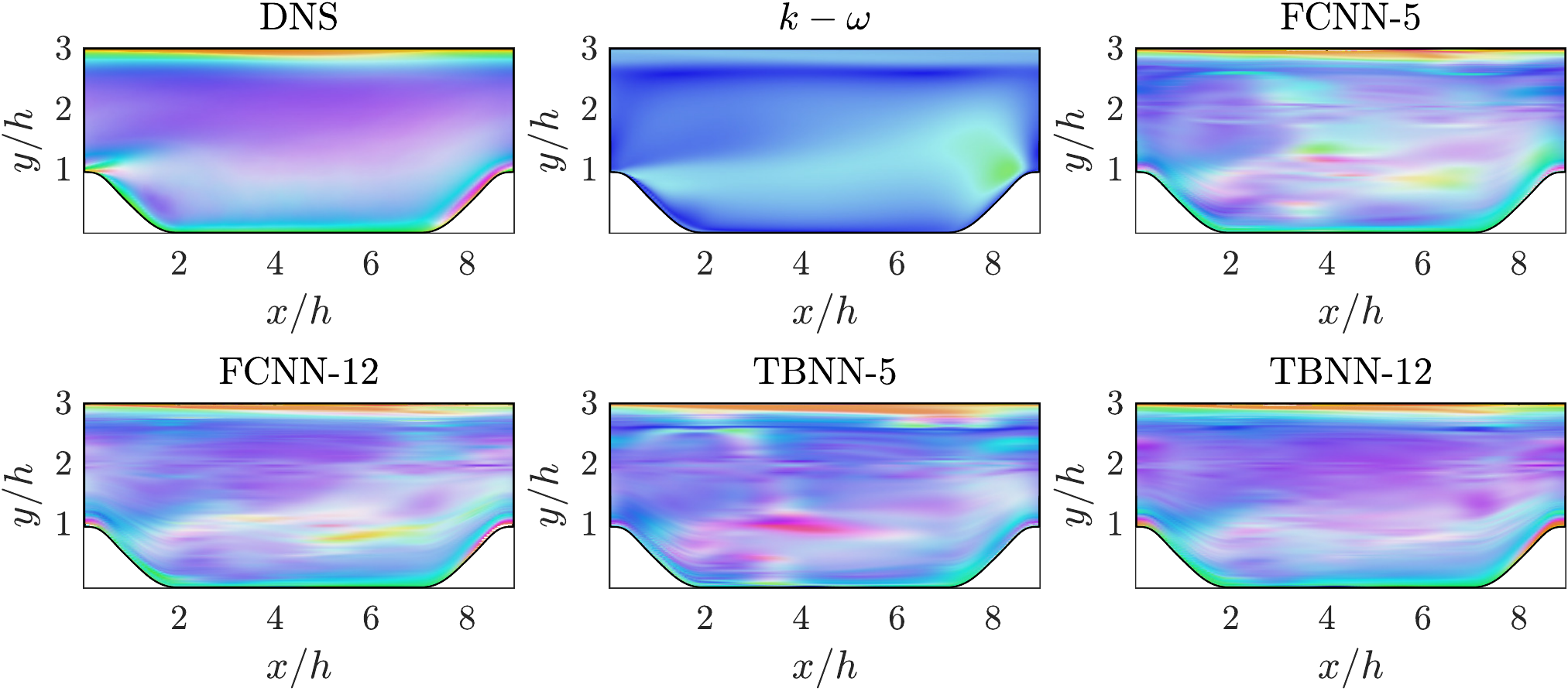}
	\caption{Turbulent states of different turbulent models in PH flow case visualized with the RGB colormap in \cref{piclumtri}.}
    \label{picstateph}
\end{figure}

\cref{picstateph} represents the turbulent states of different turbulent models in PH flow at $Re = 5600$. The contours are colored by the RGB colormap shown in \cref{piclumtri}. It can be seen that there exists 1-component turbulence (red) along the upper wall and 2-component turbulence (green) close to the upper and lower walls. At $x/h = 8$, there exists a strong 1-component turbulence, which is named the splatting effect \citep{frohlich2005highly}. 

The $k-\omega$ turbulent model cannot predict any turbulent state precisely, and it remains a plane strain turbulent state. In the splatting effect region, the large-scale eddies generated in the shear layer are convected downstream onto the upward slope, causing a high level of fluctuations in the z-direction. This also indicates that eddy viscosity models, which do not take into account such transport effects, will not be able to capture high anisotropy due to splatting. Thus, the machine learning-based turbulent model can predict these near-wall turbulent states with better precision than the $k-\omega$ model. There also exists the continuity drawback for the machine learning based turbulent model, but the TBNN model is smoother than the FCNN model, due to the FCNN-12 and TBNN-12 model execute predictions with form constraints. 

\begin{figure}[htbp]
	\centering
	\includegraphics[width=.7\textwidth]{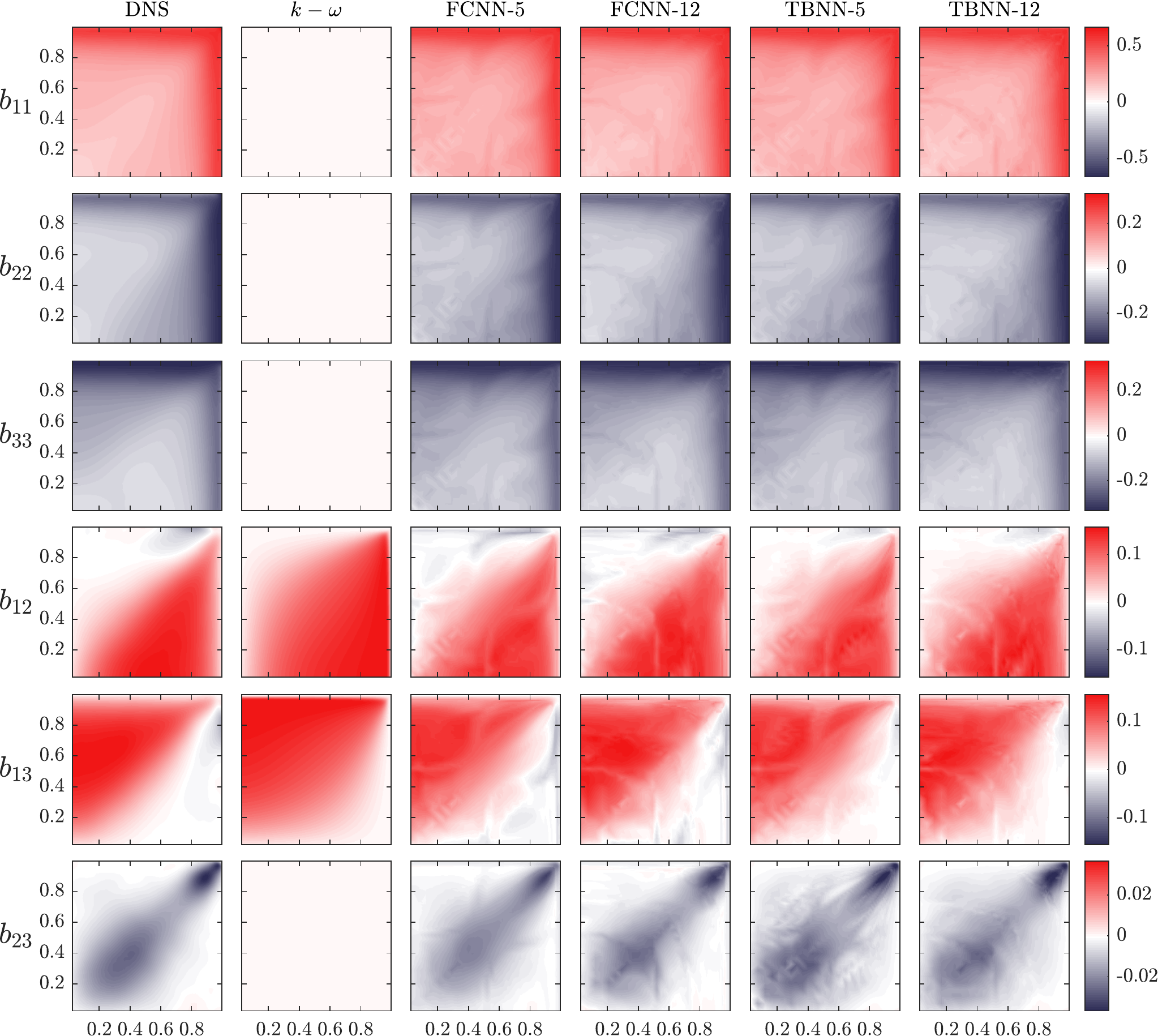}
	\caption{$b_{ij}$ prediction values of SD at $Re = 3500$. $k-\omega$ data are the RANS turbulent model prediction values.}
	\label{picsdbij}
\end{figure}

\cref{picsdbij} represents the $b_{ij}$ prediction values of the square duct flow case at $Re = 3500$. All six components of $b_{ij}$ are nonzero. The $k-\omega$ turbulent model predicts zero values in $b_{11}$, $b_{22}$, $b_{33}$ and $b_{23}$ since the RANS simulation does not yield any velocities in the $y$ direction and $z$ direction, and the velocity in the $x$ direction is fully developed with $\frac{\partial \bar{u}}{\partial x} = 0$.

\begin{figure}[!htbp]
	\centering
	\includegraphics[width=.95\textwidth]{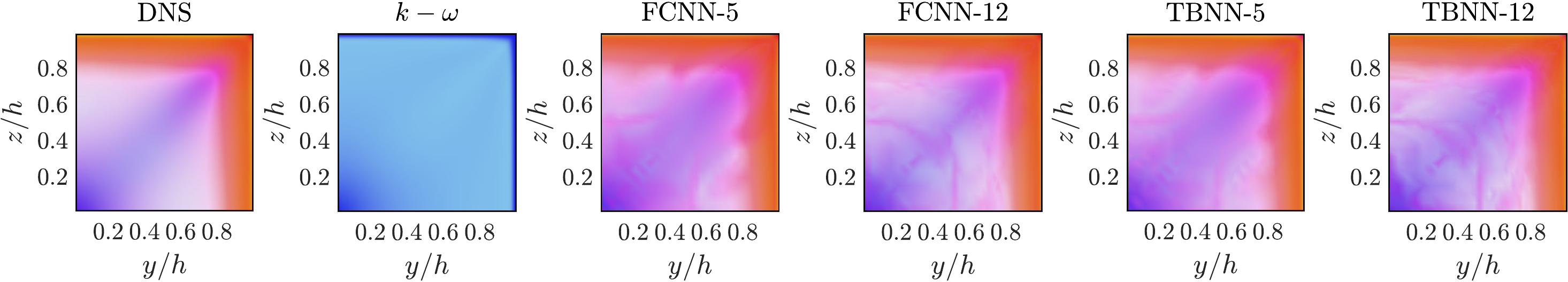}
	\caption{Turbulent states of different turbulent models in the SD flow case visualized with the RGB colormap in \cref{piclumtri}.}
	\label{picstateSD}
\end{figure}

{\blueblack
\cref{RMSEbijSD} displays the RMSE values of $b_{ij}$ prediction values of different models compared with DNS data. Similar to the periodic hill flow case, the model with $\nabla k$ will produce a better prediction of $b_{ij}$. \cref{picstateSD} represents the turbulent states of different turbulent models in the SD flow case at $Re = 3500$. It can be seen that in most regions, the turbulent state presents a mix of 1-component turbulence and 3-component turbulence. The velocity fluctuation in the $x$ direction $u$ dominates the turbulence of the 1-component turbulent state. Within the diagonal region of the flow case, the 3-component turbulent state is dominant with interaction with velocity in the $y$ and $z$ directions. The machine learning-based turbulent model can predict the 1-component turbulent well of the near wall. The machine learning turbulent model prediction values lack continuity when modeling a mixed turbulent state. However, the $k-\omega$ turbulent model cannot produce a precise turbulent state. }

\begin{table}[!htbp]
\centering
\caption{RMSE values of $b_{ij}$ predictions of different models on the square duct case.}
\label{RMSEbijSD}
\resizebox{.55\textwidth}{!}{%
\begin{tabular}{@{}cccccc@{}}
\toprule
 & \multicolumn{5}{c}{Model} \\ \cmidrule(l){2-6} 
RMSE Value & RANS & FCNN-5 & FCNN-12 & TBNN-5 & TBNN-12 \\ \midrule
$b_{11}$ & 0.4280 & 0.0226 & 0.0129 & 0.0350 & 0.0302 \\
$b_{12}$ & 0.5000 & 0.0142 & 0.0117 & 0.0145 & 0.0126 \\
$b_{13}$ & 0.0488 & 0.0144 & 0.0117 & 0.0140 & 0.0120 \\
$b_{22}$ & 0.2190 & 0.0142 & 0.0087 & 0.0186 & 0.0122 \\
$b_{23}$ & 0.0127 & 0.0030 & 0.0027 & 0.0020 & 0.0016 \\
$b_{33}$ & 0.2191 & 0.0144 & 0.0087 & 0.0185 & 0.0122 \\ \midrule
$b_{ij}$ & 0.1792 & 0.0139 & 0.0098 & 0.0174 & 0.0142 \\ \bottomrule
\end{tabular}%
}
\end{table}

\subsection{Flow velocity profile predictions}
\begin{figure}[!htbp]
	\centering
	\includegraphics[width=.98\textwidth]{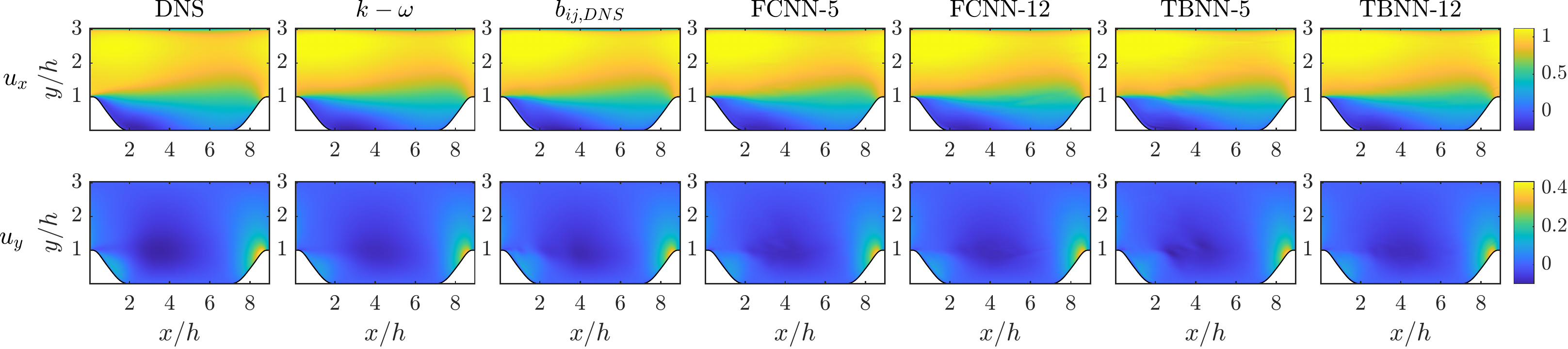}
	\caption{Predicted flow velocity distribution in the periodic hill flow case. The $b_{ij,DNS}$ data are obtained by propagating the $b_{ij}$ of the DNS data.}
	\label{picflowphall}
\end{figure}

{\blueblack
The flow fields obtained by propagating the Reynolds stress anisotropy tensor (as stated above) are displayed. The flow velocity profile based on $b_{ij,DNS}$ is obtained by propagating the Reynolds stress anisotropy tensor from high-fidelity data. It is the optimal prediction outcome achievable with the machine learning based turbulence model proposed in the present study. \cref{picflowphall} represents the flow velocity distribution in the periodic hill flow case at $Re = 5600$. The velocity distribution of the RANS model is close to that of the DNS result even with inaccurate turbulent state prediction. The velocity profile with machine learning-based $b_{ij}$ only slightly increases the accuracy in the flow field prediction.}

{\blueblack 
\cref{picflowsdvel} represents the velocity prediction result in square duct flow cases. The FCNN-5 and TBNN-5 results are omitted here. The $b_{ij,DNS}$ is obtained by propagating the Reynolds stress anisotropy tensor from DNS data. The results demonstrate that the machine learning turbulence models can predict the secondary flow phenomenon caused by Reynolds anisotropy stress. 
Meanwhile, due to the relatively discontinuous $b_{ij}$ predictions, the flow fields predicted by machine learning models show lower continuity compared to those based on DNS Reynolds shear stress tensor but not as pronounced as $b_{ij}$ prediction. \cref{picflowSD} represents the in-plane mean velocity magnitude $\sqrt{u_y^2+u_z^2}$ ($u_y$ represents the velocity component in the $y$ direction, coordinate of square duct flow case is shown in \cref{picflowcase}), which indicates the magnitude of the secondary flow of the square duct. It has proven difficult to predict square duct secondary flow using conventional turbulence models. Here one section $y/h = 0.50 $ in the square duct flow case at $Re = 3500$ is represented. The DNS data, and the data by propagating DNS $b_{ij}$ and predicted $b_{ij}$ are shown. 

 \begin{figure}[!htbp]
	\centering
	\includegraphics[width=.99\textwidth]{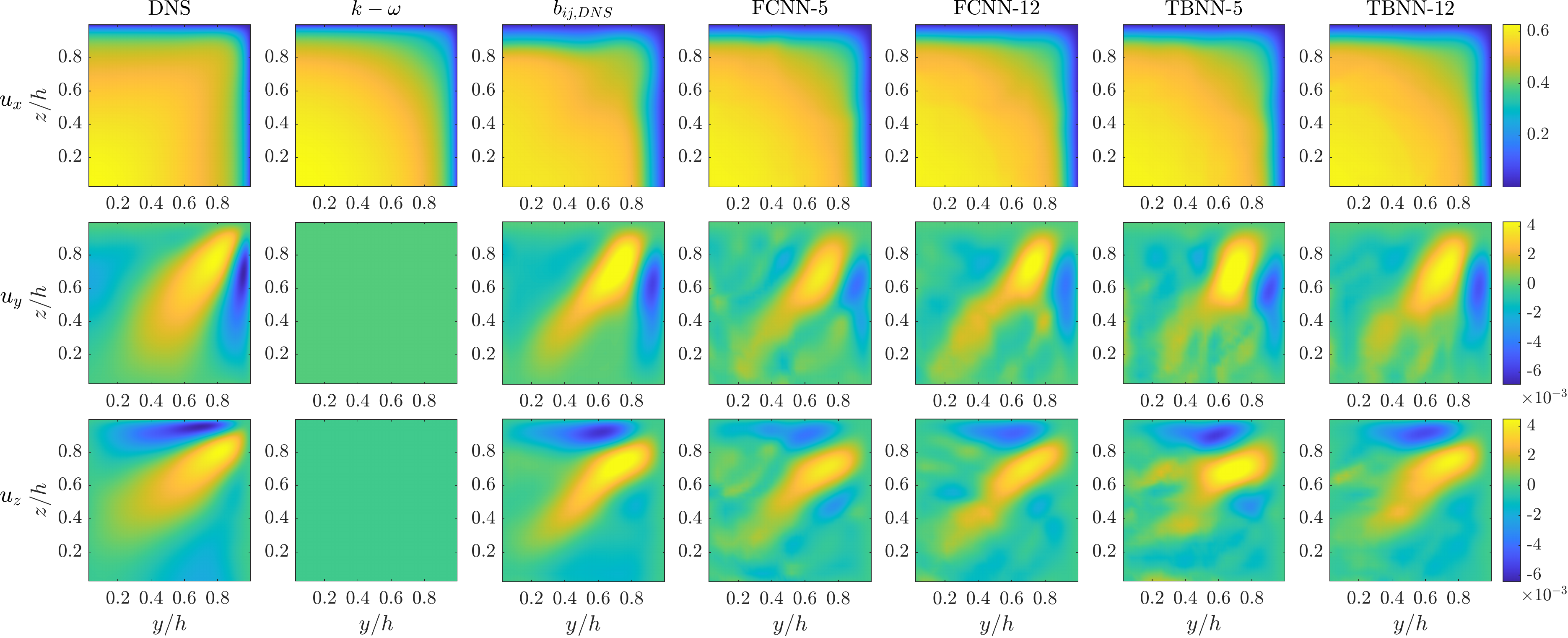}
	\caption{Predicted flow velocity profiles in the square duct flow case. The RANS values are obtained from the $k-\omega$ model. The $b_{ij,DNS}$ data are obtained by propagating the $b_{ij}$ of the DNS data.}
	\label{picflowsdvel}
\end{figure} 

Furthermore, the results from the quadratic eddy viscosity model \citep{shih1993realizable} and cubic eddy viscosity model \citep{lien1996low} are displayed. The RANS data based on $k-\omega$ turbulent model are omitted here since the magnitude remains zero. Compared with periodic hill flow case, Reynolds stress anisotropy has a bigger impact on the square cavity flow. The Reynolds anisotropy stress is a significant factor for secondary flow \cite{pinelli2010reynolds}. The machine learning based turbulent model can predict the component of $b_{ij}$ in the flow direction ($x$ direction) which results in a better velocity profile prediction. 

}

{\blueblack 
The DNS mean velocity profile is roughly reproduced by propagating $b_{ij,DNS}$ but there still exists some difference. The peak value and location are predicted well. The near-wall region ($z/h\approx 1$) matches well with the DNS data, and the location of the peak value in region $z/h\in [0.3,0.8]$ is incorrectly predicted with an error of $0.10$. The worst prediction occurs near $z/h=0$. The TBNN-12 model achieves relatively better prediction than FCNN12, with reduced amplitude in the machine learning-based turbulent model predictions. While FCNN can produce $b_{ij}$ predictions with reduced RMSE, the TBNN model achieved a slightly accurate prediction in the region $z/h\in [0.3,0.8]$. Therefore, the form constraint with prior physical or mathematics knowledge can improve the prediction of flow velocity profiles. 

 \begin{figure}[!htbp]
	\centering
	\includegraphics[width=.8\textwidth]{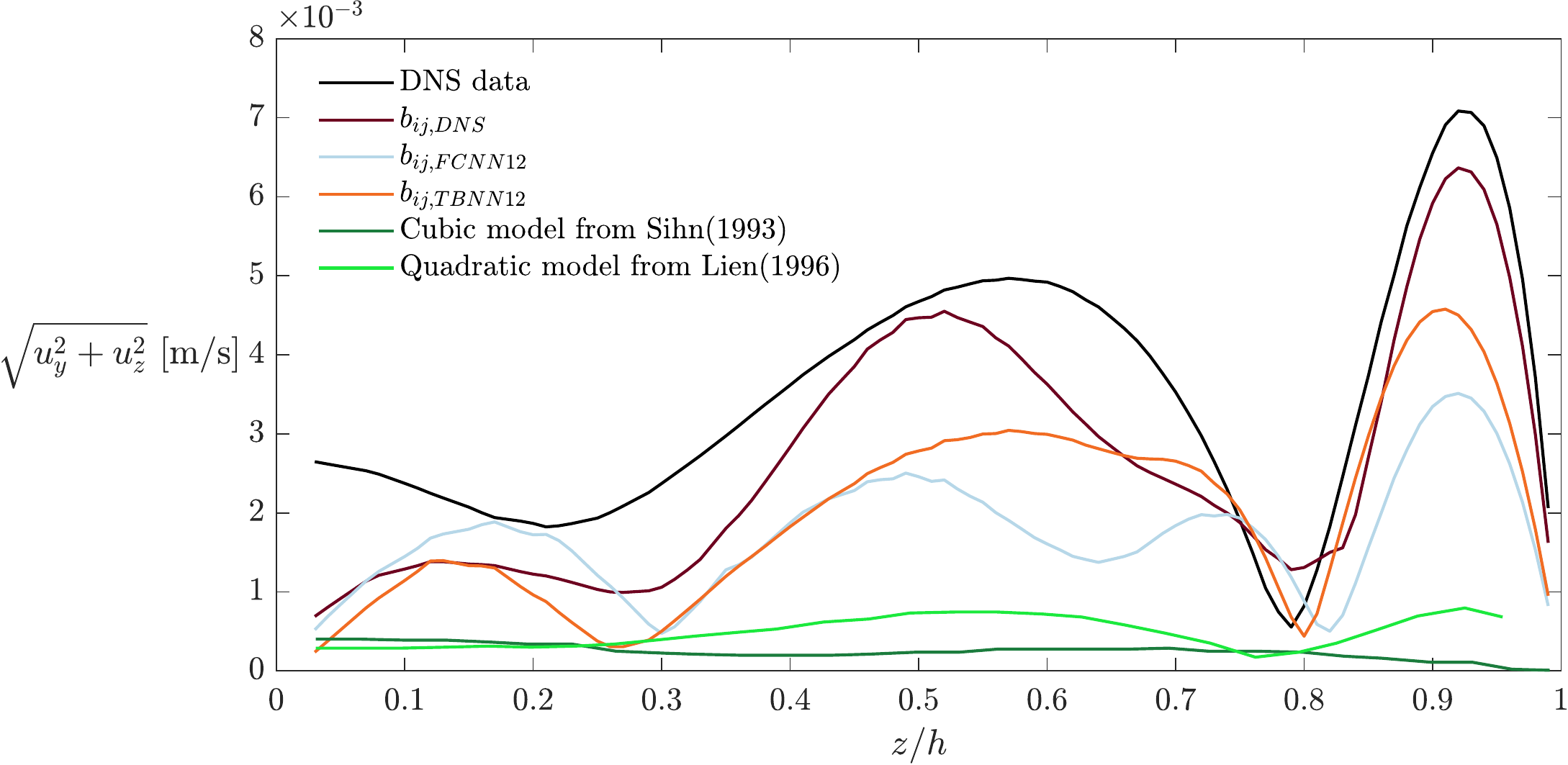}
	\caption{In-plane mean velocity profiles at section $y/h = 0.50$ of the square duct at $Re = 3500$. The cubic and quadratic models are nonlinear eddy viscosity models in \citet{shih1993realizable} and \citet{lien1996low}.}
	\label{picflowSD}
\end{figure} 

For the nonlinear eddy viscosity turbulent model, the prediction peak values of the in-plane velocity magnitude are quite poor. The quadratic turbulent model in \citet{shih1993realizable} can produce relatively correct flow velocity trends but with insufficient amplitude. Meanwhile, the prediction values given by the cubic turbulent model in \citet{lien1996low} are generally incorrect. Compared with the nonlinear eddy viscosity model, the machine learning-based turbulent model predicts a much more accurate velocity profile. Nevertheless, the $b_{ij}$ profile is highly sensitive to the velocity profile, and even a small change in $b_{ij}$ can result in a different velocity profile, such as is seen in the TBNN-12 and FCNN-12 models. The ultimate goal of machine learning based turbulent model is to implement turbulence simulation with high generalization capability, and appropriate feature selection is continuously being researched. Future studies should aim to enhance the continuity of the turbulence model based on machine learning. \citet{schmelzer2020discovery} and \citet{weatheritt2016novel} proposed different symbolic regression methods to develop an explicit formula to enhance the continuity and numerical stability. However, these methods have only been applied to establish mapping $\{\boldsymbol{S},\boldsymbol{R}\}\mapsto \boldsymbol{b}$. A more accurate turbulent model based on more invariants will be investigated in the future.}

\section{Conclusion}\label{conclu}    
In the present study, a comparative study of Reynolds stress tensor prediction methods based on invariance preservation is studied. Two kinds of machine learning structures, FCNN and TBNN, are investigated. Two invariant input sets based on ${\boldsymbol{S},\boldsymbol{R}}$ and ${\boldsymbol{S},\boldsymbol{R},\nabla k}$ are considered. The models are trained based on RANS flow fields ($k-\omega$) for input and the anisotropy tensor from high-fidelity data for output. Tensor basis neural network \cite{ling2016reynolds} and fully connected neural network are applied for training. Machine learning based on invariants ensures Galilean invariance of the turbulent model. 

{\blueblack 
The prediction of the Reynolds stress anisotropy tensor $b_{ij}$ and flow profile by propagating $b_{ij}$ are investigated in the present study. Improvement is observed with respect to the baseline simulations with the $k-\omega$ model. The FCNN model obtains a better prediction of $b_{ij}$ but shows less accuracy in the flow velocity profile compared with TBNN model. The TBNN model involves a form of constraint based on prior knowledge and improves the turbulent states. The inclusion of $\nabla k$ can increase the prediction accuracy of the Reynolds stress anisotropy tensor and flow velocity profile in comparison to the invariants defined based on $\boldsymbol{S}$, $\boldsymbol {R}$. Compared with the traditional nonlinear turbulent model \cite{shih1993realizable,lien1996low}, the machine learning-based turbulent model can improve the Reynolds stress anisotropy tensor and flow velocity profile prediction.

The drawback of the present study is that the continuity of $b_{ij}$ are relatively poor. Research on the explicit symbolic regression method \cite{schmelzer2020discovery,weatheritt2016novel} can be further applied for the invariant set based turbulent model, different from the implicit gradient decent-based MLP model as used here. Such a modification may improve the numerical stability and accuracy of the present work. Some different neural network structure considering neighborhood influence such as CNN \citep{fang2021high} will be investigated further. Meanwhile, it can be observed that the flow velocity distribution obtained based on the high-fidelity Reynolds stress still deviates from DNS results. Studies about model-consistent training \citep{duraisamy2021perspectives} such as loop training \citep{zhao2020rans} and training by indirect observation data \citep{zhang2022ensemble} will be investigated in the future to obtain the turbulent model more suitable for RANS simulation. Meanwhile, the results demonstrate considering the turbulent energy gradient can increase turbulence prediction accuracy, but this leads to an excess of independent variables. Hence, feature selection research \cite{duraisamy2015new} will also be carried out for future work.
}

\section*{Data availability}
{\blueblack 
The code and data for the neural network training are available in a public Github repository: \url{github.com/xuepengfu/tbnn_pytorch}. The code is developed based on Pytorch.}

\section*{Acknowledgments}
The authors gratefully acknowledge the financial support from National Natural Science Foundation of China under Grant Number of 52088102, 52001208, 51909159 and 52111530135, Joint Funds of the National Natural Science Foundation of China under Grant Number of U19B2013, National Science Fund for Distinguished Young Scholars under Grant Number of 51825903, State Key Laboratory of Ocean Engineering (Shanghai Jiao Tong University) under Grant Number of GKZD010081, Shenlan Project under Grant Number of SL2020PT102 and SL2021MS018, Young Elite Scientists Sponsorship Program under Grant Number of 2020QNRC001. In addition, the authors would like to express their gratitude to Dr. Mikael Kaandorp, who shared the highly resolved flow field data collected previously.

\appendix

\section{Scalar invariants of symmetric and antisymmetric tensors}\label{appenxa}
For any second order tensor $P_{ij} \in \mathbb{R}^{3\times 3}$, we can deliver a symmetric tensor $S_{ij}$ and antisymmetric tensor $W_{ij}$ as:
\begin{align}
S_{ij} &= \frac12(P_{ij}+P_{ji}),\\
W_{ij} &= \frac12(P_{ij}-P_{ji}).
\end{align}

\noindent For a symmetric tensor, a simpler form based on the principal axis system can be expressed as:
\begin{equation}
    S_{ij}=\begin{bmatrix}
  e_1& 0 &0 \\
 0 &  e_2&0 \\
 0 & 0 & e_3
\end{bmatrix},
\end{equation}
where $e_i$ is the principal components of $S_{ij}$ in the direction of the principal axis. And the antisymmetric tensor can be expressed as:
\begin{equation}
    W_{i j}=\dfrac12 \left[\begin{array}{ccc}
0 &  \omega_{3} &  \omega_{2} \\
-\omega_{3} & 0 &  \omega_{1} \\
- \omega_{2} & - \omega_{1} & 0
\end{array}\right]. 
\end{equation}

There are three main invariants of $S_{ij}$ expressed as:
\begin{equation}
\begin{array}{l}
J_{1}=e_{1}+e_{2}+e_{3},\\
J_{2}=e_{1}^{2}+e_{2}^{2}+e_{3}^{2},\\
J_{3}=e_{1}^{3}+e_{2}^{3}+e_{3}^{3}.
\end{array}
\end{equation}

\noindent Additionally, widely used invariants called principle invariants are expressed as follows:
\begin{equation}
\begin{array}{l}
I_{1}=\operatorname{tr}(S_{ij})=e_{1}+e_{2}+e_{3}, \\
I_{2}=\frac{1}{2}\left((\operatorname{tr}(S_{ij}))^{2}-\operatorname{tr}\left(S_{ij}^{2}\right)\right)=e_{1} e_{2}+e_{1} e_{3}+e_{2} e_{3}, \\
I_{3}=\operatorname{det}(S_{ij})=e_{1} e_{2} e_{3}.
\end{array}    
\end{equation}

\noindent The relationship between $I_i$ and $J_i$ is:
\begin{equation}
\begin{array}{l}
J_{1}=I_{1}, \\
J_{2}=I_{1}^{2}-2 I_{2}, \\
J_{3}=I_{1}^{3}-3 I_{1} I_{2}+3 I_{3}.
\end{array}   
\end{equation}

In \cite{zheng1994theory,eringen1980mechanics,johnson2016handbook}, they all state that there exist seven invariants based on symmetric and antisymmetric tensors:
\begin{equation}
\begin{array}{c}
    \lambda_1 = tr(S_{ij}),\; \lambda_2 = tr(S_{ij}^2),\; \lambda_3 = tr(S_{ij}^3),\\
 \lambda_4 = tr(W_{ij}^2),\; \lambda_5 = tr(S_{ij}W_{ij}^2),\; \lambda_6 = tr(S_{ij}^2W_{ij}^2),\; \lambda_7 = tr(S_{ij}^2W_{ij}^2S_{ij}W_{ij}).
\end{array}
\end{equation}

\noindent However, \citet{pope1975more} only used the first six invariants for turbulent modeling. Pope and the reference \cite{spencer1958theory} have not clarified it, which induced some reference mistakes \cite{berrone2022invariances1}. In the following, we will prove that $\lambda_7^2$ can be expressed as a function of the first six invariants.

Following the above representation of $S_{ij}$ and $W_{ij}$, we can have the invariants $\lambda_1 = J_1$, $\lambda_2 = J_2$ and $\lambda_3 = J_3$. $\lambda_4$ can be expressed as:
\begin{equation}
    \lambda_4 = tr(W_{ij}^2) =-\frac12 (w_1^2 + w_2^2 + w_3^2),
\end{equation}
and $\lambda_5$ is:
\begin{align}
\lambda_5 =tr(S_{ij}W_{ij}^2)= -\frac14( e_1w_2^2 + e_2w_1^2 + e_1w_3^2 + e_3w_1^2 + e_2w_3^2 + e_3w_2^2).
\end{align}    

\noindent We assume that $J_1=0$ for representation simplification (this conclusion also holds when $J_1\neq 0$), which can deliver:
\begin{equation}
\lambda_5 = \frac14(e_1w_1^2 + e_2w_2^2 + e_3w_3^2).    
\end{equation}

\noindent The $\lambda_6$ can be expressed as:
\begin{equation}
    \lambda_6 = tr(S_{ij}^2W_{ij}^2) = -\frac14 (e_1^2w_2^2 + e_2^2w_1^2 + e_1^2w_3^2 + e_3^2w_1^2 + e_2^2w_3^2 + e_3^2w_2^2).
\end{equation}

\noindent Considering the main invariants, the $\lambda_6$ can be simplified as:
\begin{equation}
    \lambda_6 = -\frac14\left[(2J_2-J_1)(w_1^2+w_2^2+w_3^2)-(e_1^2w_1^2+e_2^2w_2^2+e_3^2w_3^2)\right],
\end{equation}
and the last invariants $\lambda_7$ can be expressed as:
\begin{align}
\lambda_7  = tr(S_{ij}^2W_{ij}^2S_{ij}W_{ij}) & = \frac18w_1w_2w_3\left(e_1^2e_2 - e_1^2e_3 - e_1e_2^2 + e_1e_3^2 + e_2^2e_3 -e_2e_3^2\right)\\
&=\frac18(e_2-e_1)(e_3-e_2)(e_1-e_3)w_1w_2w_3.
\end{align}

\noindent We can see that $\lambda_7$ is the function based on $w_i$. It is easy to find three invariants based on $w_i^2$ from $\lambda_4$, $\lambda_5$ and $\lambda_6$:
\begin{align}
K_1 &= w_1^2+w_2^2+w_3^2,\\
K_2 &= e_1w_1^2 + e_2w_2^2 + e_3w_3^2,\\
K_3 &= e_1^2w_1^2+e_2^2w_2^2+e_3^2w_3^2.
\end{align}
\noindent This equation can be solved as follows:
\begin{align}
w_1^2 & = \frac{(K_3 - K_2e_2 - K_2e_3 + K_1e_2e_3)}{(e_1 - e_2)(e_1 - e_3)},\\
w_2^2 & = \frac{(K_3 - K_2e_1 - K_2e_3 + K_1e_1e_3)}{(e_2 - e_1)(e_2 - e_3)},\\
w_3^2 & = \frac{(K_3 - K_2e_1 - K_2e_2 + K_1e_1e_2)}{(e_1 - e_3)(e_2 - e_3)}.
\end{align}
\noindent Considering $J_1=0$, $w_i$ can be solved as: 
\begin{align}
w_1^2 & = \frac{(K_3 + K_2e_1  + K_1e_2e_3)}{(e_1 - e_2)(e_1 - e_3)},\\
w_2^2 & = \frac{(K_3 + K_2e_2  + K_1e_1e_3)}{(e_2 - e_1)(e_2 - e_3)},\\
w_3^2 & = \frac{(K_3 + K_2e_3  + K_1e_1e_2)}{(e_1 - e_3)(e_2 - e_3)}.
\end{align}

\noindent Thus, $\lambda_7^2$ can be expressed as:
\begin{align}
\lambda_7^2  = \left[tr(S_{ij}^2W_{ij}^2S_{ij}W_{ij})\right]^2 & = \frac{1}{64}(e_2-e_1)^2(e_3-e_2)^2(e_1-e_3)^2w_1^2w_2^2w_3^2\\ 
& = \frac{1}{128}[2K_1^3I_3+2K_1K_2I_3J_3+2K_1^2K_3I_3J_1\\
&\quad +K_1K_2^2(J_2^2-J_1J_3-I_3J_2+J_2I_2)\\
&\quad +2K_1K_2K_3(J_1J_2-J_3)+2K_1K_3^2I_2+K_3^2I_3\\
&\quad +2K_2^2K_3I_2+2K_2K_3^2I_1+2K_3^3],
\end{align}
which means $\lambda_7^2 = f(\lambda_1,\lambda_2,\lambda_3,\lambda_4,\lambda_5,\lambda_6)$, so the seven invariants are not independent. There only exist six independent invariants of $\lambda_1$ to $\lambda_6$. In fact, this phenomenon is called syzygy in tensor function community, which means that the invariants are related by an implicit function. In references \cite{zheng1994theory,eringen1980mechanics,johnson2016handbook}, the complete set of invariants is given, not a minimum complete set. In the training process based on tensor function theory, there may exist inappropriate data weights of different invariants. 

\section{Basis tensor and scalar function}\label{appendixB}
\begin{figure}[!htbp]
	\centering
	\includegraphics[width=.4\textwidth]{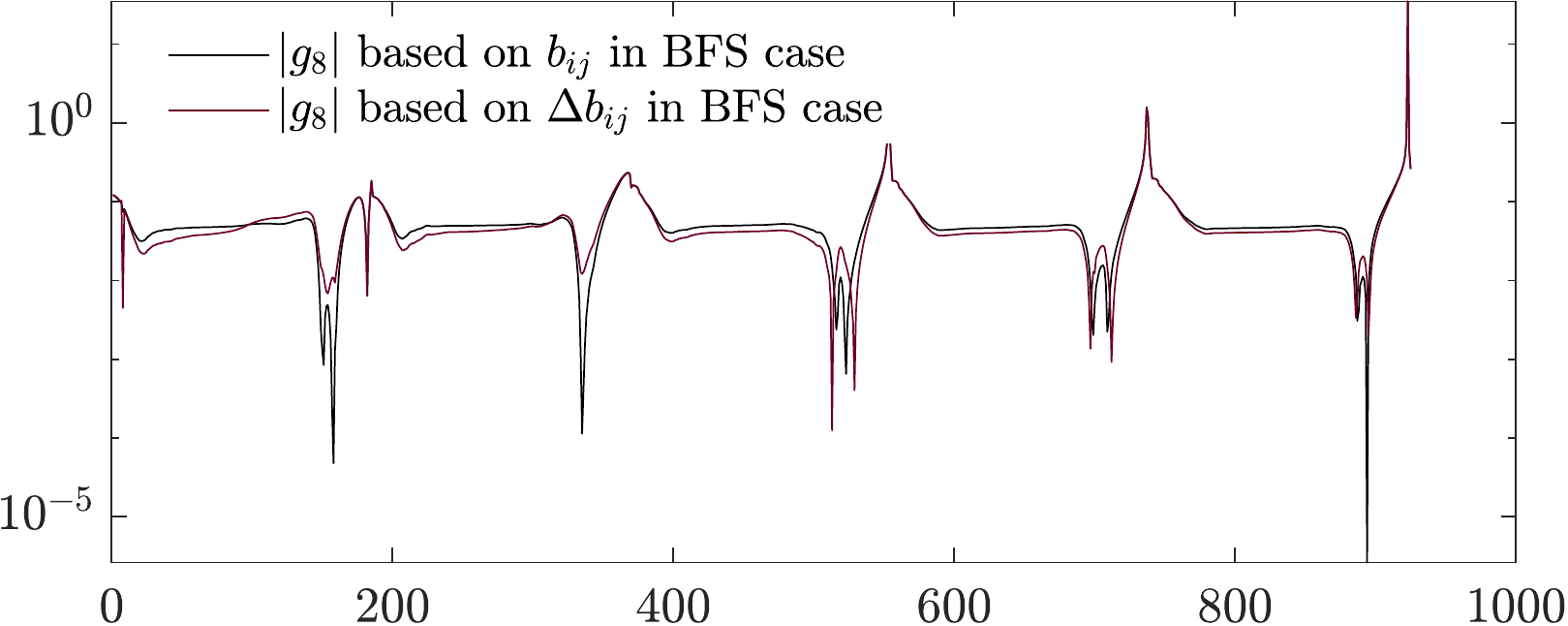}
	\caption{Absolute scalar function values $|g_8|$ based on $b_{ij}$ and $\Delta b_{ij}$ in the backward-facing step flow case. $|g_8|$ is the eighth value of $\boldsymbol{C}_{10\times1}$ obtained based on \cref{c10}.}
	\label{piccoe}
\end{figure}

\begin{figure}[!htbp]
	\centering
	\includegraphics[width=.7\textwidth]{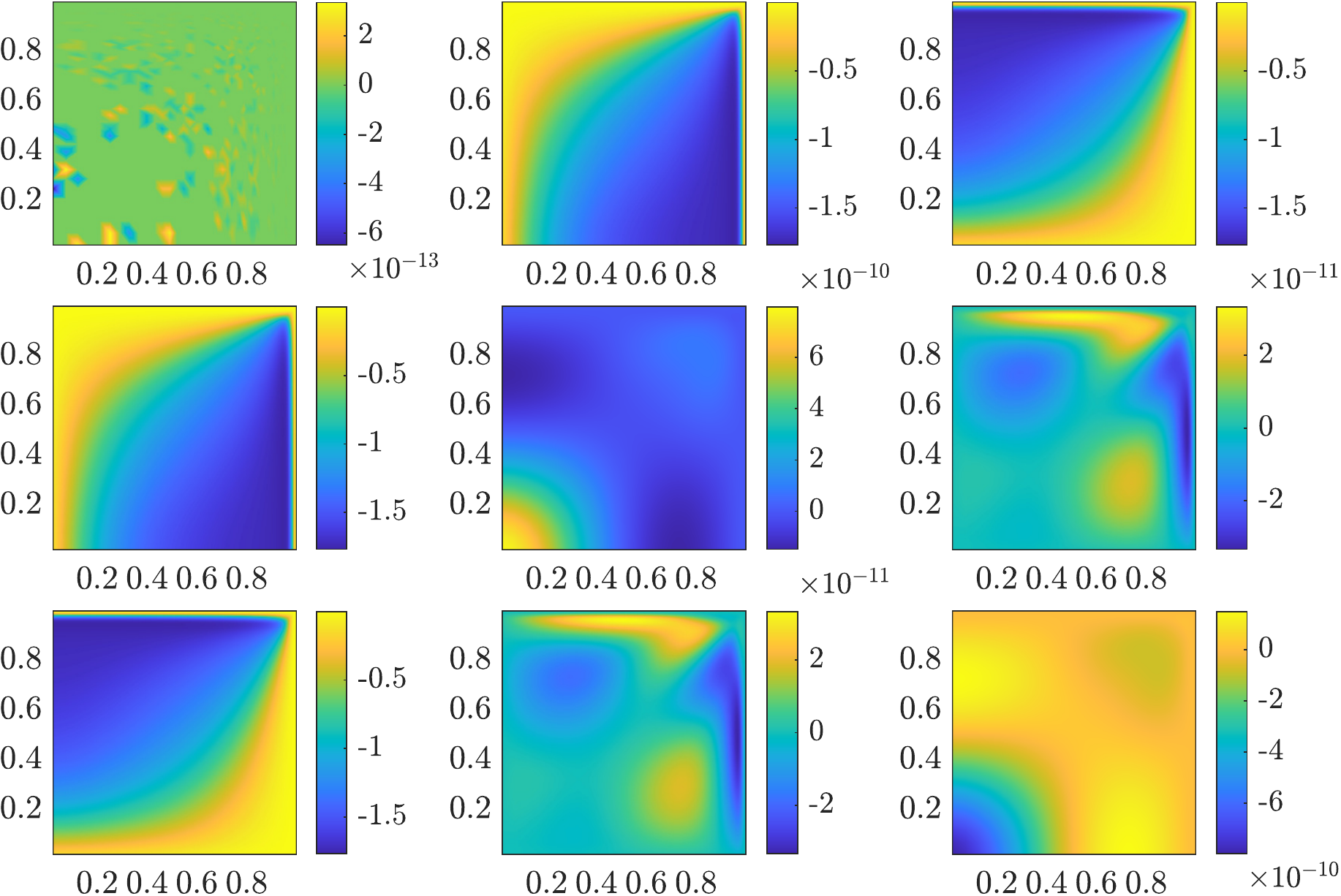}
	\caption{Visualization of tensor basis $T^{(1)}$ in the SD flow case at $Re = 3500$. $T^{(1)}$ is second order tensor with six independent value, the subplots correspond to the contours of each component.}
	\label{picT1}
\end{figure}

{\blueblack 
The scalar function $\boldsymbol{C}_{10\times1}$ obtained based on $b_{ij}$ and $\Delta b_{ij}$ are estimated and shown in this appendix. We reshape the output anisotropy tensor $\boldsymbol{b}_{3\times 3}$  and input tensor basis $\boldsymbol{B}_{3\times3\times10}$ as $\boldsymbol{b}_{9\times 1}$ and $\boldsymbol{B}_{9\times10}$, respectively. Then we have:
\begin{equation}
    \boldsymbol{B}_{9\times10} \cdot \boldsymbol{C}_{10\times1} = \boldsymbol{b}_{9\times 1},
\end{equation}
where $\boldsymbol{C}_{10\times1}$ is the scalar function value $g_i$ in \cref{popeb}. $\boldsymbol{C}_{10\times1}$ can be estimated by the least squares method as:
\begin{equation}\label{c10}
    \boldsymbol{C}_{10\times1} = (\boldsymbol{B}_{9\times10}^T\boldsymbol{B}_{9\times10})^{-1}\boldsymbol{B}_{9\times10}^T\boldsymbol{b}_{9\times 1}.
\end{equation}

\cref{piccoe} displays the absolute scalar function values $|g_8|$ based on $b_{ij}$ and $\Delta b_{ij}$ in the BFS case. It can be observed that the scalar function values can exhibit minor fluctuations based on $\Delta b_{ij}$ in comparison to $b_{ij}$. \cref{picT1} displays the contour plot of the tensor basis $T^{(1)}$ of the square duct flow case at $Re = 3500$.}

\end{document}